\documentclass[a4paper,11pt]{article}
\pdfoutput=1 

\usepackage{jcappub} 

\usepackage[utf8]{inputenc}
\usepackage{amsmath}
\usepackage{graphicx}
\usepackage{amssymb}
\usepackage{bbold}
\usepackage{hyperref}
\usepackage{lipsum}
\usepackage{booktabs}
\usepackage{tikz}

\usepackage[colorlinks=true,citecolor=blue,linkcolor=blue]{hyperref}
\usepackage[normalem]{ulem}
\usepackage{amsmath,amssymb}
\usepackage{mathtools}
\usepackage{verbatim}
\usepackage{titlesec}               
\usepackage{epsfig}
\usepackage{graphicx}               
\usepackage{url}
\usepackage{color}
\usepackage{multirow}
\usepackage{placeins}
\usepackage[dvipsnames]{xcolor}
\usepackage{epstopdf}
\usepackage{siunitx}
\usepackage{fontawesome}
\usepackage{enumitem}
\usepackage[capitalize]{cleveref}
\usepackage{lipsum}
\usepackage{gensymb}
\usepackage{booktabs}               
\usepackage{xspace}                 
\usepackage{pifont}
\usepackage{marvosym }
\usepackage{tikz-feynman}
\usepackage{tikz} 
\usetikzlibrary{shapes,arrows}
\usetikzlibrary{decorations.pathmorphing,decorations.markings}
\usetikzlibrary{snakes}
\usepackage{orcidlink}  
\allowdisplaybreaks

\usepackage{bbm}
\usepackage{slashed}

\hypersetup{colorlinks=true,allcolors=[rgb]{0.5,0.0,0.5}}

\usepackage{natbib}

\definecolor{customgreen}{HTML}{7F170E}
\definecolor{customred}{HTML}{EA3323}

\begin{document}


\title{Vanishing Acts: Quantifying Black Hole Formation with the DSNB Signal}

\author{T. Charissé \orcidlink{0009-0005-7329-9845}} 
\emailAdd{tchariss@students.uni-mainz.de}
\author{D. Maksimovi\'c \orcidlink{0000-0002-7387-293X}}
\emailAdd{damaksim@uni-mainz.de}
\author{G. A. Parker \orcidlink{0009-0000-1836-8696}}
\emailAdd{geparker@uni-mainz.de}
\author{M. Wurm \orcidlink{0000-0003-2711-0915}}
\emailAdd{michael.wurm@uni-mainz.de}
\affiliation{PRISMA${}^{+}$ Cluster of Excellence and Institut f{\"u}r Physik,
Johannes Gutenberg-Universit{\"a}t Mainz,
55099 Mainz, Germany}


\begin{abstract}
    {The diffuse supernova neutrino background (DSNB) created by stellar core-collapses throughout cosmic history is on the verge of discovery, with SK-Gd showing early deviations from the background expectation and JUNO starting to take data. However, the interpretation of early DSNB data will face significant challenges due to degeneracies between astrophysical parameters and uncertainties in supernova neutrino modeling. We explore how complementary astronomical observations can break these degeneracies and, in this context, we investigate whether early DSNB observations can constrain invisible supernovae, which have no optical emission but are powerful neutrino sources before being swallowed by a forming black hole. Leveraging the differences in the spectra between invisible and visible supernovae, we estimate the sensitivity of 1) detecting the existence of invisible supernovae, and 2) determining the fraction of invisible supernovae. Finally, we discuss how these conclusions depend on the spectral parameters of the black hole-forming component.}
\end{abstract}
\maketitle

\section{Introduction}

The Diffuse Supernova Neutrino Background (DSNB) are all neutrinos produced by core-collapse supernovae throughout the Universe, offering a way to study the entire population of core-collapse events \cite{Nakazato2013,Lunardini2012,Horiuchi2017,Nakazato2015}. Steady progress is being made in the technological development towards its discovery. The Super-Kamiokande experiment has added gadolinium to enhance signal sensitivity and background suppression, and the data has already begun to show a deviation from the pure-background expectation \cite{Harada2024DSNB}. At the time of writing, the JUNO experiment has just started physics data taking \cite{JUNOpnpp, Abusleme2022}. The expected rate for the combination of both of these large, sensitive detectors is $\sim5$ events detected per year. With expected signal-to-background ratios close to unity, a positive detection of the DSNB seems only a matter of time. So as we draw closer to a discovery of the DSNB signal, we ask \textit{What will we see?} and \textit{What can we learn?} \cite{MITP2024DSNB} based on the first years of DSNB data.

Previous studies of the DSNB signal \cite{Mathews2014, Moller2018,Kresse2021} have shown a high level of degeneracy between the relevant physics entering the calculation. 

If, beyond its mere discovery, the DSNB is to be used as a tool to study properties of the underlying supernovae, we must first address the large astrophysical uncertainties, including the overall normalization of the supernova rate, its dependence on progenitor mass, $M$, and on the redshift $z$. Our incomplete knowledge of stellar populations also enters.

It is unlikely that these degeneracies can be broken at the signal level based on the low expected DSNB detection rates. Under these circumstances, the only path forward is to imform the interpretation of neurino spectra using results from other astronomical observations. In the past, the stellar collapse rate has been inferred by observing galaxies of different redshifts and extrapolating the star formation history \cite{Hopkins2006}. This approach is increasingly replaced by direct observations of the $z$-dependent rate of visible SN explosions. Sky surveys like the Legacy Survey of Space and Time (LSST) at the Vera Rubin Telescope \cite{LSST} will see out to redshift $z\lesssim1$, revealing the origins of $\sim90\%$ of the expected 10–26 MeV DSNB event rate \cite{Lien2010}. On the other hand, gravitational wave observations of neutron star and black hole mergers will further improve our understanding of the maximum masses and the equation of state of neutron stars \cite{Ecker2025}, thus shrinking the uncertainty margin on the predictions for the neutrino spectra of visible supernovae. 

The present paper explores whether the combination of these advancements with the early DSNB signal can be exploited to address one of the last big unknowns, which is the fraction of \textit{invisible} supernovae. By definition not accessible to optical astronomy, black hole-forming SNe are powerful emitters of neutrinos before being extinguished by the emerging black hole. Here, we study under which circumstances the invisible supernova neutrino component becomes significant in the DSNB signal.

The paper is structured as follows: we begin with the physics that enters the calculation of the DSNB flux in \cref{sec:theory}, emphasizing the cosmic core-collapse rate, and neutrino spectrum in \cref{sec:SN_rate,sec:nu_spec}, respectively. In \cref{sec:exp}, we discuss the relevant detectors for the DSNB and their backgrounds in detail. In \cref{sec:method}, we describe our analysis method and present the results in \cref{sec:res}. Next, \cref{sec:discuss} discusses the dependence on the analysis and the underlying model. We conclude in \cref{sec:conc}.
\section{Modelling the DSNB}
\label{sec:theory}
The isotropic DSNB flux, for neutrino flavour $\beta$ as a function of neutrino energy at Earth, $E$, is defined as:
\begin{equation}
\Phi_{\nu_\beta}(E) =c \int_0^{z_{\max }} \frac{1}{H(z)} \times \mathcal{R}_{\mathrm{CCSN}}(z) \times F_{\nu_\beta}[(1+z)E] \ d z
\end{equation}

where $c$ is the speed of light, and we set maximum redshift $z_{\max }$ to 5. There are three parts to this redshift integral: 

\paragraph{1) $\frac{1}{H(z)}:$} Here, $H(z)$ is the Hubble parameter as a function of redshift. This accounts for the expansion history of the Universe. Recall that within standard $\Lambda$CDM cosmology $H(z) = H_0\sqrt{\Omega_m(1+z)^3+\Omega_{\Lambda}}$, where $H_0$ is the Hubble constant, and $\Omega_m=0.3$, $\Omega_{\Lambda}=0.7$ are the fractions of the cosmic energy density in matter and dark energy, respectively.

\paragraph{2) $\mathcal{R}_{\mathrm{CCSN}}(z):$} The \textit{cosmic core-collapse rate} is the rate of collapse events per unit of comoving volume. As stellar lifetimes are a \textit{brief candle} \cite{macbeth} on cosmic timescales, this is closely linked to the star-formation history of the Universe. This component contains an integral over progenitor mass, $M$.

\paragraph{3) $F_{\nu_\beta}[(1+z)E]:$} Here, $F$ is the time-integrated average supernova neutrino spectrum, and the energy $(1+z)E$ accounts for a neutrino which is emitted at redshift $z$. The population-average is achieved by weighting with the initial mass function (IMF).\\

We do not vary the cosmological component, but the astrophysics \textbf{2)} and neutrino physics components \textbf{3)} of the DSNB will be discussed in the following sections.
\subsection{Cosmic core-collapse rate}
\label{sec:SN_rate}
Throughout this work, we follow the astrophysical setup described in Kresse \textit{et al}, and interested readers should consult \cite{Kresse2021} and references therein. We reproduce here some specific details for illustration. The cosmic core-collapse rate can be described as follows, where $\psi_*(z)$ describes the cosmic star-formation history \cite{Yksel2008}, $\phi$ is the IMF \cite{Baldry2003}, and $M$ is the progenitor mass. 
$$
R_{\mathrm{CC}}(z)=\psi_*(z) \frac{\int_{8.7 \mathrm{M}_{\odot}}^{125 \mathrm{M}_{\odot}} \mathrm{d} M \phi(M)}{\int_{0.1 \mathrm{M}_{\odot}}^{125 \mathrm{M}} \mathrm{d}M M \phi(M)} .
$$

\subsection{Neutrino Spectrum}
\label{sec:nu_spec}
The neutrino spectrum per unit time is well-approximated by a pinched-thermal function \cite{Keil2003, Tamborra2012}. 

\begin{equation}
\varphi^0(E) = \left(\frac{L}{\langle E\rangle}\right)\frac{(\alpha+1)^{(\alpha+1)}}{\Gamma(\alpha+1)}\left(\frac{E}{\langle E \rangle}\right)^\alpha \exp \left(-\frac{(\alpha+1)E}{\langle E \rangle}\right).
\label{pinch}
\end{equation}
where $L$ is the neutrino luminosity, $\langle E \rangle$ is the mean energy, and $\langle E^2 \rangle$ is the second moment of energy. The pinching parameter, $\alpha$, which quantifies the spectral pinching, which is the deviation from a thermal spectrum caused by neutrinos scattering as they diffuse outwards \cite{Raffelt2001}. A thermal spectrum (Fermi-Dirac distribution with zero chemical potential) corresponds to \cref{pinch} when $\alpha = 2.3$ \cite{Keil2003, Tamborra2012}.

In \cref{pinch}, we define $\varphi^0(E)$ which does not account for the Mikheyev-Smirnov-Wolfenstein (MSW) effect, a resonant flavour conversion due to the supernova density profile \cite{Wolfenstein1978-wy, Wolfenstein1979-cc, MS1, MS2, Dighe2000-rv}. Since we consider only the inverse beta decay (IBD) channel in this work, $\bar{\nu}_e + p \rightarrow e^{+} + n$, we present only the analytic approximation for the MSW effect of electron antineutrinos in normal ordering (which will be considered for the rest of the work), where $\theta_{12}$ is the solar mixing angle:
\begin{equation}
    \varphi_{\bar{\nu}_e} = \varphi^0_{\bar{\nu}_e}{\cos^2\theta_{12} + \varphi^0_{\nu_x}\sin^2\theta_{12}}.
\end{equation}
In this work, we use the sophisticated DSNB simulations described in \cite{Kresse2021}, which are publicly available at \cite{Garching}. These models rely on 200 one-dimensional simulations of non-rotating single stars, which explode and produce a neutrino signal via a \textit{'neutrino engine'} \cite{Ertl2016}. By calibrating the neutrino engine with different strengths, this generates a population with a different fraction of black hole-forming collapses, astrophysically consistent across the supernova mass range. The strongest neutrino engine leads to the smallest black hole fraction. 
Note that we disregard the $\sim 5\%$ of supernovae in the data that correspond to electron-capture supernova which occur in the transition between white dwarf formation and iron core-collapse supernovae.

\begin{table}[h!]
\centering
\begin{tabular}{lccccc}
\toprule
Engine Model & S19.8 & N20 & W18 & W15 & W20 \\
\midrule
$f_{\mathrm{BH}}$ & 17.8\% & 22.8\% & 26.9\% & 29.1\% & 41.7\% \\
 & 
\begin{tikzpicture}[scale=0.3]
  \fill[customgreen] (0,0) -- (0,1) arc (90:154.04:1) -- cycle;
  \fill[customred] (0,0) -- (154.04:1) arc (154.04:450:1) -- cycle;
\end{tikzpicture} &
\begin{tikzpicture}[scale=0.3]
  \fill[customgreen] (0,0) -- (0,1) arc (90:172.08:1) -- cycle;
  \fill[customred] (0,0) -- (172.08:1) arc (172.08:450:1) -- cycle;
\end{tikzpicture} &
\begin{tikzpicture}[scale=0.3]
  \fill[customgreen] (0,0) -- (0,1) arc (90:186.84:1) -- cycle;
  \fill[customred] (0,0) -- (186.84:1) arc (186.84:450:1) -- cycle;
\end{tikzpicture} &
\begin{tikzpicture}[scale=0.3]
  \fill[customgreen] (0,0) -- (0,1) arc (90:194.76:1) -- cycle;
  \fill[customred] (0,0) -- (194.76:1) arc (194.76:450:1) -- cycle;
\end{tikzpicture} &
\begin{tikzpicture}[scale=0.3]
  \fill[customgreen] (0,0) -- (0,1) arc (90:240.12:1) -- cycle;
  \fill[customred] (0,0) -- (240.12:1) arc (240.12:450:1) -- cycle;
\end{tikzpicture} \\
\bottomrule
\end{tabular}

\caption{Black hole fraction for the five models, which have a neutrino engine calibrated on progenitor Z9.6 by A. Heger \cite{Kresse2021} and the progenitors S19.8, N20, W18, W15, and W20, respectively \cite{Sukhbold2016}.}
\label{tab:pie}
\end{table}

\begin{figure}[h!]
    \centering
    \includegraphics[width=0.8\linewidth]{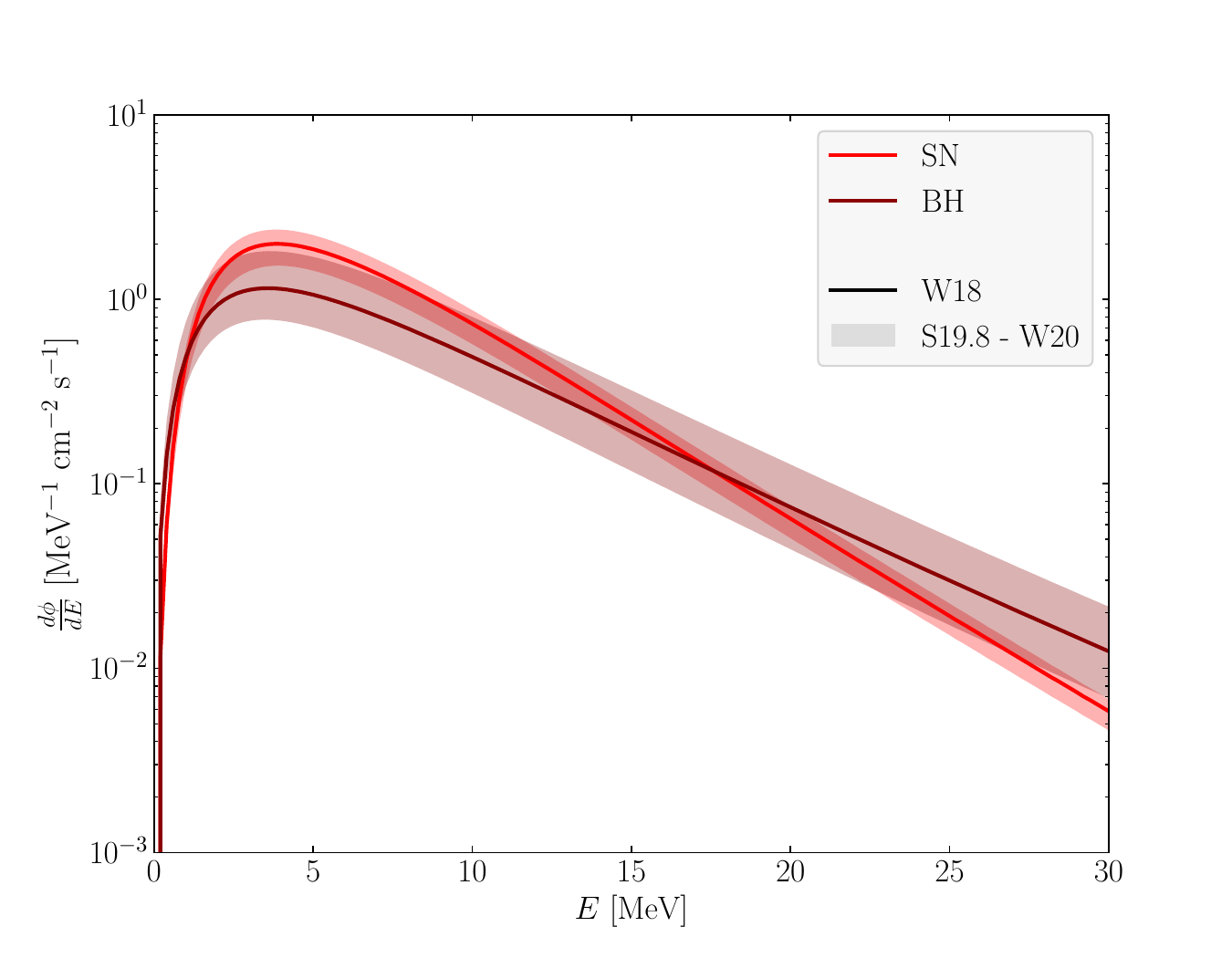}
    \caption{Plot of the DSNB flux spectrum for the median fiducial model (W18), with bands for extreme models (S19.8 - W20) for both neutron star and black hole forming core-collapse supernovae.}
    \label{fig:flux1}
\end{figure}

For each supernova engine, S19.8, N20, W18, W15, and W20 \cite{Sukhbold2016, Kresse2021}, see \cref{tab:pie,fig:flux1}, the event rates are calculated for both black hole-forming core-collapses, and neutron star-forming collapses. 
 
The resulting neutrino spectra have different characteristics, such as the black hole-forming collapses having a harder spectrum because of the large increase in luminosities and average energies of the neutrinos emitted before black hole formation \cite{Sumiyoshi2006}. This key difference in overall normalization and spectral shape is what is leveraged in this study to access \textit{invisible} stellar deaths. We approximate the visible component as having a neutron star-forming spectrum and the invisible as a black hole-forming spectrum. This could be further complicated by core-collapse events which are obscured by dust or astrophysical structures resulting in being both optically invisible but contributing to the DSNB with a neutron star-forming spectrum, for example. In contrast, there could be black hole-forming collapses that still yield supernova explosions, being visible events that contribute to the DSNB with a black hole-forming spectra. However, we expect both of these contributions to be negligible~\cite{Hans-Thomas-Janka, Burrows2025}. 
\section{Experimental setup and backgrounds}
\label{sec:exp}
For the present study, we mostly concentrate on the signals that will be collected in the two large-scale experiments that will take data relevant for the DSNB search during the upcoming years: Super-Kamiokande and JUNO. Here, we shortly list our assumptions made on detector dimensions and operation times. We also explain the background levels and spectra that are crucial to understand DSNB detection in both experiments. For the sake of this study, we treat the up-coming Hyper-Kamiokande (without gadolinium) in a simplified fashion that will also be outlined below. In all cases, we have been referring to publications and information given by the respective experimental collaborations. Note that in all three cases we rely on substantially simplified assumptions about the experiments that cannot reflect the more careful work done by the respective collaborations, using full detector simulations and background data samples.

\begin{figure}[ht!]
    \centering
    \includegraphics[height=0.3\textheight]{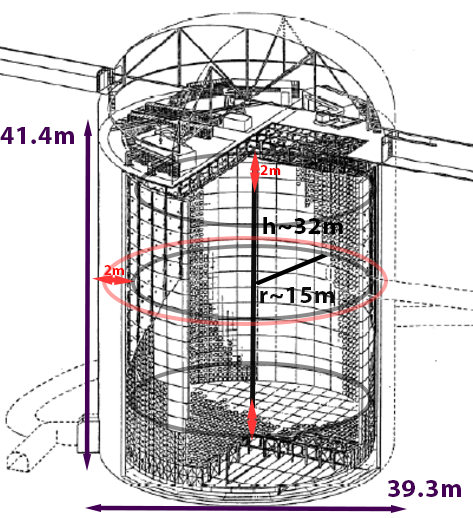}
    \hspace{0.1\textwidth}
    \includegraphics[height=0.3\textheight]{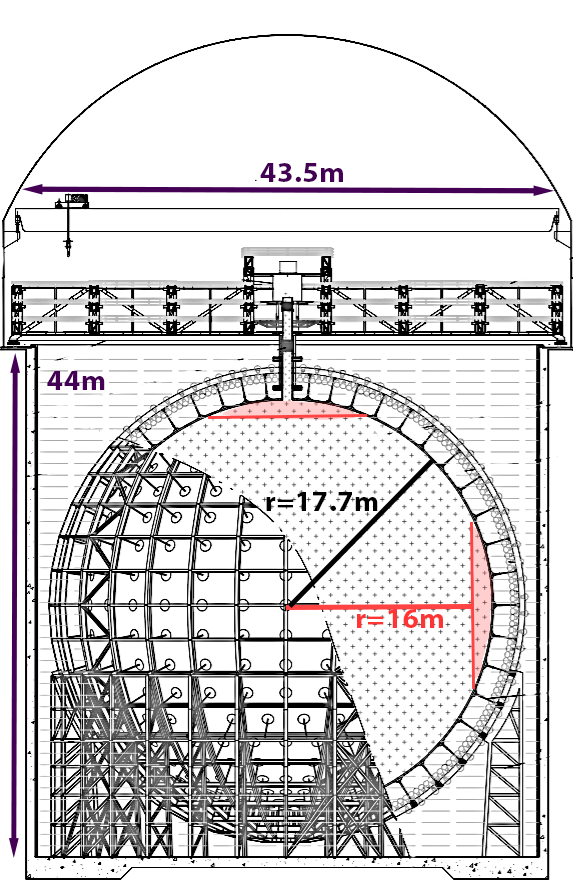}
    \caption{Detector layouts of SK (\textbf{Left}) and JUNO (\textbf{Right}). The red parts shown are exclusion areas, to mitigate backgrounds such as fast neutrons from spallation processes in the surrounding rock. The areas excluded for SK are $\sim$2 m from the PMTs, leading to a cylindrical fiducial volume of $22.5$\,kt with $r\sim15$ m and $h\sim 32$ m \cite{SuperKSRNS2012}. For JUNO, the top part with a vertical position $z>16$\,m and the sides with $r=\sqrt{Z^{2}+{r^2}_{XY}}$ with $r_{XY} > 16$\,m are excluded \cite{Abusleme2022}.}
    \label{fig:detectors}
\end{figure}
\footnotetext[1]{The quoted detection efficiency \cite{Kunxian:2015ymr} is slightly lower than the current neutron detection efficiency of 75\% which corresponds to a $\mathrm{Gd}_2 (\mathrm{SO}_4)_3$ loading fraction of 0.079\%. Before mid-2022, the fraction of $\mathrm{Gd}_2 (\mathrm{SO}_4)_3$ was at 0.02\%, resulting in lower neutron capture efficiency of $\sim$50\% \cite{Abe2022}.}
\footnotetext[2]{Note that the quoted loss in signal efficiency for JUNO includes Triple Coincidence and Muon Veto, but is in fact further reduced through Pulse Shape Discrimination from \cite{Abusleme2022}, which is energy-dependent.}

\begin{table}[h!]
    \centering
    \begin{tabular}{l|ccc}
        Experiment & JUNO & SK-Gd & HK  \\
        \hline
        Fiducial mass (kt)          & 17 & 22.5   & 45  \\
        Number of free protons      & 1.45$\times$10$^{33}$  & 1.5$\times$10$^{33}$  & 1.2$\times$10$^{34}$  \\
        IBD detection efficiency    & 100\%  & 67\%\footnotemark[1]  & 20\%  \\ 
        Signal efficiency & 87.6\%\footnotemark[2] & 92\% & 92\% \\
        Background level (per kt$\times$yr) & 0.027  & 0.077  & 0.154  \\
        Start of operation          & 2025  & 2020  & 2027  \\
    \end{tabular}
    \caption{The detector and operation parameters assumed for JUNO, Super-Kamiokande with gadolinium-loading (SK-Gd) and Hyper-Kamiokande (HK) throughout this paper. These simplified assumptions are meant to capture the general characteristics of each experiment \cite{SuperKSRNS2012,Abusleme2022,JUNOpnpp,HyperKDesign2018}.}
    \label{tab:det_par}
\end{table}

\subsection{Experiments}

\paragraph{JUNO} The Jiangmen Underground Neutrino Observatory (JUNO) is based on a 20\,kt liquid scintillator detector. With its primary goal being the spectroscopy of reactor antineutrino oscillations and the unprecedented size of the scintillator target, JUNO is expected to provide excellent sensitivity to the inverse beta decays (IBDs) on hydrogen (protons) induced by the electron antineutrino component of the DSNB. Crucially, neutrons from IBDs are captured on hydrogen in the liquid scintillator. The 2.2\,MeV gamma released by this process causes a bright scintillation signal that can be detected with high efficiency (close to unity), providing a clear coincidence signature to suppress single-event backgrounds. The signal efficiency is slightly reduced due to vetoes introduced for cosmic muon followers and suppression of triple coincidences \cite{Abusleme2022}. As described below, there is a sizable correlated background arising from the Neutral-Current (NC) interactions of atmospheric neutrinos in the scintillator. Laboratory and simulation studies suggest that this background can be largely suppressed by machine learning-based pulse-shape discrimination (PSD).  
The corresponding background levels taken from Ref.~\cite{Abusleme2022} in the nominal detection window from 12 to 30\,MeV are displayed in the left panel of \cref{fig:bg-spectra}. JUNO has started scintillator filling in February 2025, physics data taking with the full detector is expected to commence in the second half of 2025.

\paragraph{Super-Kamiokande (SK)} SK is a water Cherenkov detector with a large fiducial mass of 22.5\,kt for the DSNB. As in JUNO, the DSNB is detected via IBDs on free protons. In 2020, the detector performance was enhanced by the addition of gadolinium-sulfate to the water. The upgraded SK-Gd features increased detection efficiency for IBDs due to the 8\,MeV gamma cascade induced by neutron capture on gadolinium to $\sim$75\% (formerly, only 20\% of the IBDs with neutron captures on hydrogen could be identified). The slight loss in signal efficiency is mostly due to a cut on the Cherenkov angle of the prompt event. At the Neutrino 2024 conference, the collaboration released the most recent set of intermediary SK-Gd results on the event spectra after background rejection \cite{Harada2024DSNB}. These results are reproduced by the right panel of \cref{fig:bg-spectra}. The dominant background arises from charged-current (CC) interactions of atmospheric neutrinos that create ''invisible muons'' below Cherenkov result and accompanied by a single detected neutron removed from an oxygen nucleus. The DSNB signal retention for SK is heavily influenced by neutron capture rate and background discrimination.

\paragraph{Hyper-Kamiokande (HK)} HK is currently under construction at an underground site not far from the Kamioka mine. The primary focus of the experiment is on the measurement of CP violation from the (upgraded) T2K beam. HK will be considerably larger than SK, featuring about eight times the fiducial mass for the DSNB search. However, the detector has a lower rock overburden than SK and crucially there is no Gd-loading foreseen for the initial phase of the experiment. Hence, we expect that the efficiency for coincident detection of IBD events is $\sim$20\%, similar to SK before the Gd-loading. For simplicity, we do not consider the higher background levels expected due to shallower depth. Also, we are not considering the information from DSNB single events (i.e.~without neutron tag). HK is scheduled to begin operation in 2027.

\subsection{Backgrounds}
\begin{figure}[h!]
    \centering
    \includegraphics[width=0.5\linewidth]{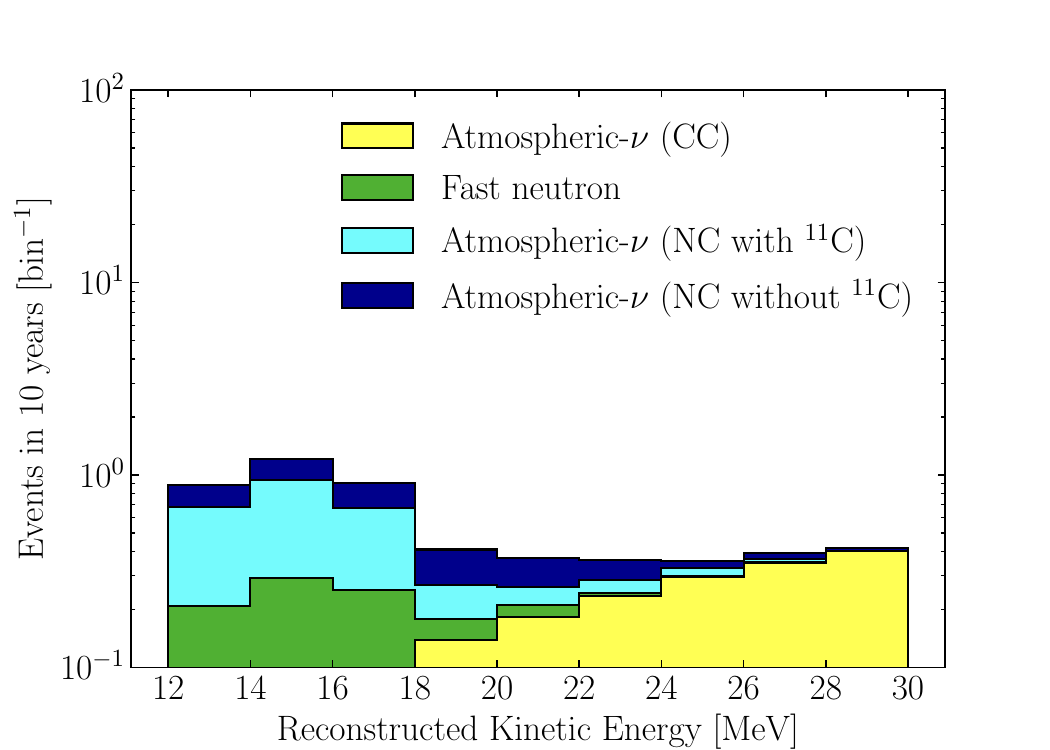}
    \hspace{-0.5cm}
    \includegraphics[width=0.5\linewidth]{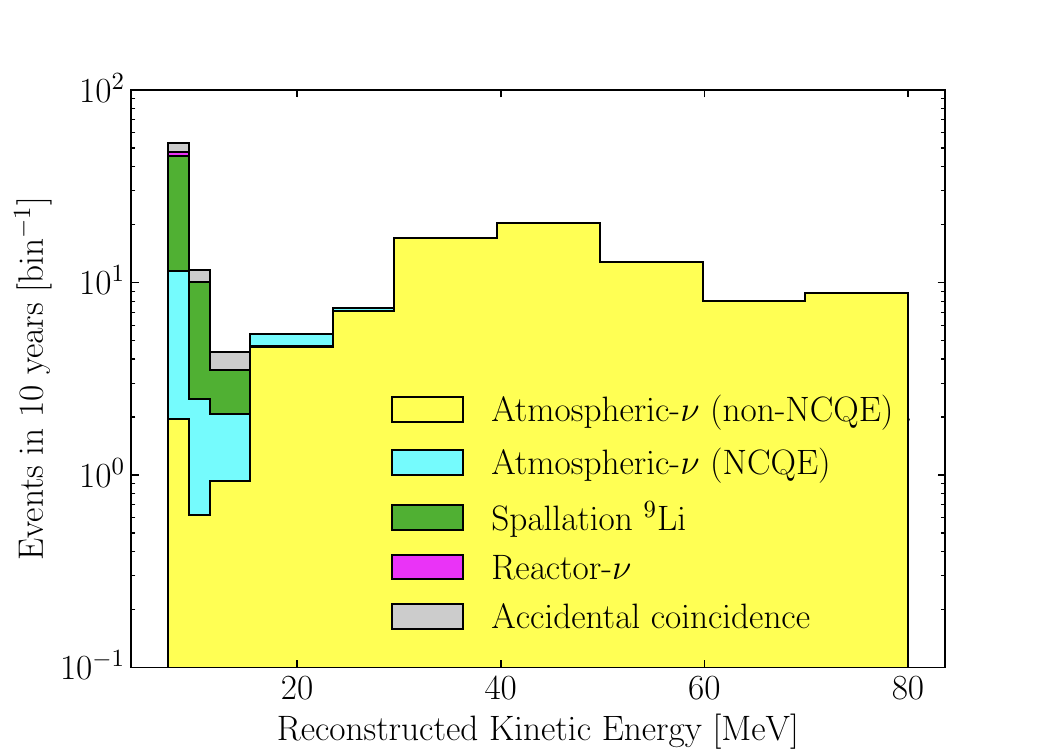}
    \caption{Background spectra assumed for JUNO  (\textbf{Left}) and SK-Gd  (\textbf{Right}). JUNO's prediction is taken from Ref.~\cite{Abusleme2022} and is restricted to the DSNB observation window, while the background spectra shown for SK-Gd are based on Ref.~\cite{Harada2024DSNB} and include as well large sideband areas. In both plots, we group similar backgrounds with similar colors.} 
    \label{fig:bg-spectra}
\end{figure}

The relevant background sources for JUNO (left panel) or SK-Gd/HK, respectively, (right panel) are shown in \cref{fig:bg-spectra}. Commonly, the observation window is defined based on the energy of the prompt event and ranges from about 10 to 30 MeV. Here, we review the most relevant background sources.

\paragraph{Reactor antineutrinos} The ample flux of $\bar\nu_e$'s emitted by nuclear power plants cannot be discriminated on an event-by-event level from the IBDs induced by DSNB neutrinos. For all detectors, this ample background effectively prevents the detection in the energy range below 10 - 12\,MeV. The exact threshold depends on the experimental site and the energy resolution of the detector. 

\paragraph{Spallation background ($^9$Li)} Cosmic muons traversing the detector target can induce spallation reactions on carbon resp.~oxygen, thereby producing radioactive isotopes. There are two specific isotopes, $^8$He and $^9$Li, that decay with half-lives of $\sim$200\,ms via $\beta n$-emission, mimicking the IBD signal. The spectral endpoint for the more relevant $^9$Li decays is towards the lower end of the observation window. Several experiments have explored possibilities for vetoing this background based on time and spatial coincidences with the preceding parent muons and neutrons \cite{Nairat2025} (see \cite{CosmogenicInducedSpallationRemoval} for a recent study). While important for reactor $\bar\nu_e$ detection in JUNO, $^9$Li is not considered here since the $\beta$-spectrum ends below the 12\,MeV detection threshold.

\paragraph{Fast neutrons} Cosmic muons traversing the rock surrounding the detector cavern can create at times high-energy neutrons. These fast neutrons have a non-zero probability to pass through the outer veto layers undetected but then induce a sizable proton recoil, followed by thermalization and capture in the detector medium. This background only appears in liquid scintillator were protons of several tens MeV in energy can create a visible signal comparable to that of IBDs. Events of this kind are, however, located mostly at the outer rim of the detection volume and thus can be largely removed by a fiducial volume cut. In turn, fast neutrons are irrelevant for SK-Gd/HK due to the lack of Cherenkov emission.
    
\paragraph{Charged-current (CC) interactions of atmospheric neutrinos} Towards higher energies, CC interactions of atmospheric neutrinos are the primary background in water Cherenkov detectors. This background comes in two variations: The low flux of atmospheric $\bar\nu_e$'s at energies comparable to the DSNB undergo IBD interactions and dominate the DSNB signal in both water and scintillator from about 30\,MeV. In addition, water Cherenkov detectors suffer from atmospheric $\bar\nu_\mu$'s slightly higher-energy that will create low-energy muons below the Cherenkov threshold in the final state. The subsequent muon decays create Cherenkov signals from the Michel electrons that can be misinterpreted as an IBD if there is in addition a knock-out neutron from the initial CC interaction. These so-called ''invisible muons'' pose a much less severe background for scintillator detectors where the final-state muon creates an additional signal.

\paragraph{Neutral-current (NC) interactions of atmospheric neutrinos} NC interactions of atmospheric neutrinos at several hundred MeV in energy will often create low-energy recoil particles in the detector target, featuring a selection of nuclear fragments, protons and gammas that in combination make up a prompt signal, while additional neutrons can provide a delayed capture signal. While featuring a relatively large rate compared to the DSNB signal, SK-Gd has been able to demonstrate that this background can be efficiently discriminated based on the washed-out Cherenkov signature of the prompt event \cite{Harada2024DSNB,Maksimovic2021,Fujita2023}. This is less straightforward in scintillator detectors where many low energy protons and hadrons will add a faint scintillation signal to the prompt event. However, preliminary studies performed for JUNO suggest that high suppression factors can be reached based on pulse-shape discrimination without substantial loss of DSNB signal efficiency\cite{Abusleme2022}. 
\section{Analysis Method}
\label{sec:method}
In this section, we describe the analysis which is split into two statistical tests. The first quantifies the sensitivity of the DSNB events to having \textbf{any} invisible supernova component. The second determines to what significance it is possible to measure what \textbf{fraction} of the total flux comes from these black hole-forming collapses. We employ a negative log-likelihood with Poisson statistics to account for the low event counts, where $P=\lambda^k \exp (-\lambda) / k!$.

All event rates are calculated from the fluxes described in \cref{sec:nu_spec} by applying the IBD cross-section \cite{Strumia2003-dx}, and detector effects including target volume and detector efficiencies. The expected backgrounds were introduced in \cref{sec:exp}. Their rates $b_j$ are taken from \cite{Abusleme2022,Harada2024DSNB} 
but scaled to our exposure. Note that in definitions we suppress all arguments except the parameter of interest, the fraction of added BH spectra, $f_{\mathrm{inv}}$, or the black hole fraction, $f_{\mathrm{BH}}$, for clarity.  This quantity is the fraction of massive stars that end their life as a black hole instead of a neutron star and is estimated by tracking stars and observing them 'disappear'. One such study, with 11 years of data at the aptly named Large Binocular Telescope, yields $f_{\textrm{BH}}=0.16^{+0.23}_{-0.12}$ (90\% CL) \cite{Neustadt2021-zj}. 

\paragraph{Test 1:} \textbf{\textit{Is there a black hole-forming component present in the DSNB signal?}}\\

With future sky surveys revealing the overall normalization of the neutron star component, and gravitational wave observations giving insight into the spectral parameters, we expect to have a clear prediction of the DSNB flux created by visible supernovae \cite{Lien2010,Ecker2025}. To detect the flux component of invisible supernovae, we use the neutrino event spectra for neutron star-forming collapses (denoted NS) and the black hole forming-collapses (BH) discussed in \cref{sec:nu_spec}. Assuming an improved knowledge of supernova astrophysics, we fixing the NS component of the DSNB signal, $s^{\mathrm{NS}}$, and vary the BH component, $s^{\mathrm{BH}}$, from $0-200\%$. The fraction of added BH spectra, $f_{\mathrm{inv}}$ ($0-2$), is \textbf{NOT} the black hole-fraction. When $f_{\mathrm{inv}}=1$, then $f_{\mathrm{BH}}$ is consistent with the particular model, and has the value given in \cref{tab:pie}. 

However, scaling $f_{\mathrm{inv}}$ is a useful statistical exercise to determine at what level of additional flux we gain sensitivity to the invisible component, for each of the five DSNB models predicting different event numbers and $f_{\mathrm{BH}}$ from \cite{Kresse2021}. We write
\begin{align}
    \chi^2(f_{\mathrm{inv}})\simeq-2\sum_{i} \ln L_i(f_{\mathrm{inv}})
    +\left(\frac{\eta}{\sigma_{\mathrm{NS}}}\right)^2+\sum_j \left(\frac{\beta_j}{\sigma_{j}}\right)^2,
\end{align}
including the likelihood
\begin{align} 
L_i(f_{\mathrm{inv}}) = P\Bigl( s^{\mathrm{NS}}_{i}+\sum_j b_{ji},\quad \eta s^{\mathrm{NS}}_{i}+\sum_j \beta_j b_{ji}+f_{\mathrm{inv}}s^{\mathrm{BH}}_{i}\Bigl),
\label{eq: likelihoodWT}
\end{align}
where $\eta$ and $\beta_j$ are the pull factors on the NS component, $s^{\mathrm{NS}}$, and backgrounds, $b_j$,  respectively. According to the  current best estimates of the experiments \cite{Harada2024DSNB, Abusleme2022}, the associated uncertainties are set to $\sigma_{\mathrm{NS}}=0.3$ and $\sigma_j = \small \begin{cases} 0.2 & \text{NC } \\[-.7ex]  0.3 & \text{non-NC} \end{cases}$ .

\paragraph{Test 2:} \textbf{\textit{What is the black hole fraction?}}

\begin{figure}[h!]
    \centering

    \includegraphics[width=0.8\linewidth]{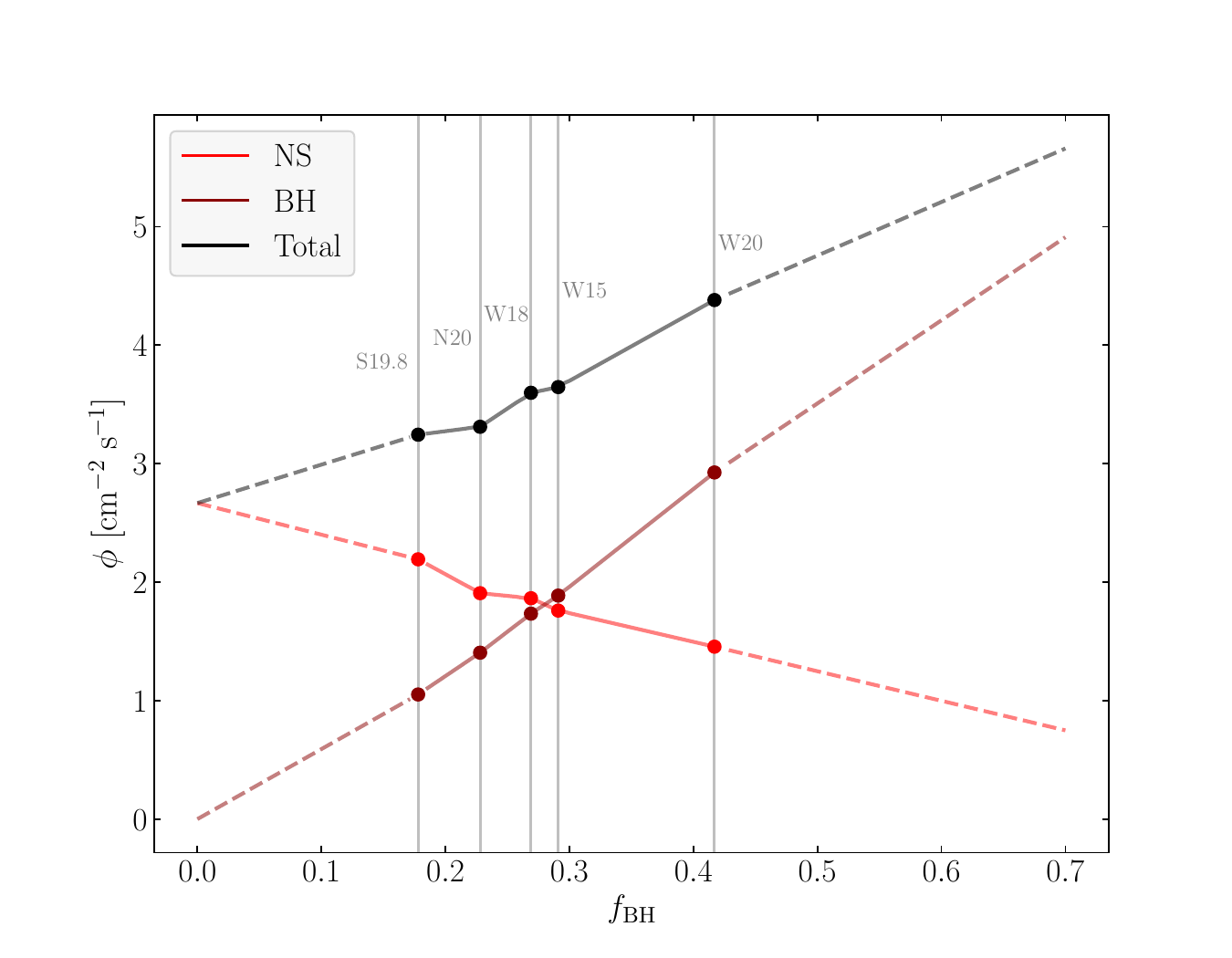}
    \caption{Neutrino flux integrated from 0 - 90 MeV as a function of black hole-fraction, $f_{\mathrm{BH}}$, for both neutron star and black hole-forming supernova. The solid lines indicate interpolation between the 'true' points (dots) for each model. Outside 0.178 and 0.417, we scale using the extreme models (dotted lines).}
    \label{fig:model1}
\end{figure}

We aim to determine the combined sensitivity of current and upcoming monolithic neutrino detectors, JUNO, SK-Gd and HK, to $f_{\mathrm{BH}}$, by letting both the visible and invisible components vary as a function of black hole fraction (i.e.~not fixing the normalization of the visible component, different from \textbf{Test 1}).

In this test, we assume that both the visible and invisible components are accurately described by the the models \cite{Garching, Kresse2021}. To estimate the sensitivity to $f_{\mathrm{BH}}$, the five DSNB models are taken as true points, between which the respective NS and BH components are interpolated, yielding a continuous set of spectra as a function of $f_{\mathrm{BH}}$ (solid lines in \cref{fig:model1}). Outside the boundaries of the models, that range from $17.8\%$ to $41.7\%$, the NS and BH components are scaled proportionally. We note that these extreme regions are not in agreement with a realistic star formation history and are primarily included for completeness (dashed lines in \cref{fig:model1}). The $\chi^2$ function used is
\begin{align}
    \chi^2(f_\mathrm{BH})\simeq-2\sum_{i} \ln L_i(f_\mathrm{BH})
    +\sum_j \left(\frac{\beta_j}{\sigma_{j}}\right)^2,
\end{align}
including the likelihood 
\begin{align} 
L_i({f_\mathrm{BH}}) = P\Bigl( s_{i}+\sum_j b_{ji},\quad S_{i}(f_\mathrm{BH})+\sum_j \beta_j b_{ji}\Bigl),
\end{align}
where $s = s^{\mathrm{NS}}+s^{\mathrm{BH}}$ is the total neutrino rate of one model, and $S(f_\mathrm{BH})$ is the total neutrino rate according to a linear interpolation between the five existing engine models, see \cref{fig:model1}. The backgrounds are denoted $b_j$, with the associated pull factors and uncertainties, $\beta_j$ and $\sigma_j$, as before. Here, we assume the existing models fully describe supernova neutrino spectra from the visible component, and therefore, do not marginalize over this $\eta$, in contrast to \textbf{Test 1}.

\section{Results}
\label{sec:res}

In this section, we discuss the results of our statistical tests. The $\chi^2$-profiles for 10 years of data with JUNO and with SK are presented, showing the effect of including spectral information, as well as the difference in sensitivity between detectors.

\paragraph{Test 1:} \textbf{\textit{Is there a black hole-forming component present in the DSNB signal?}}\\

\begin{figure}[h!]
    \centering
    \includegraphics[width=0.5\linewidth]{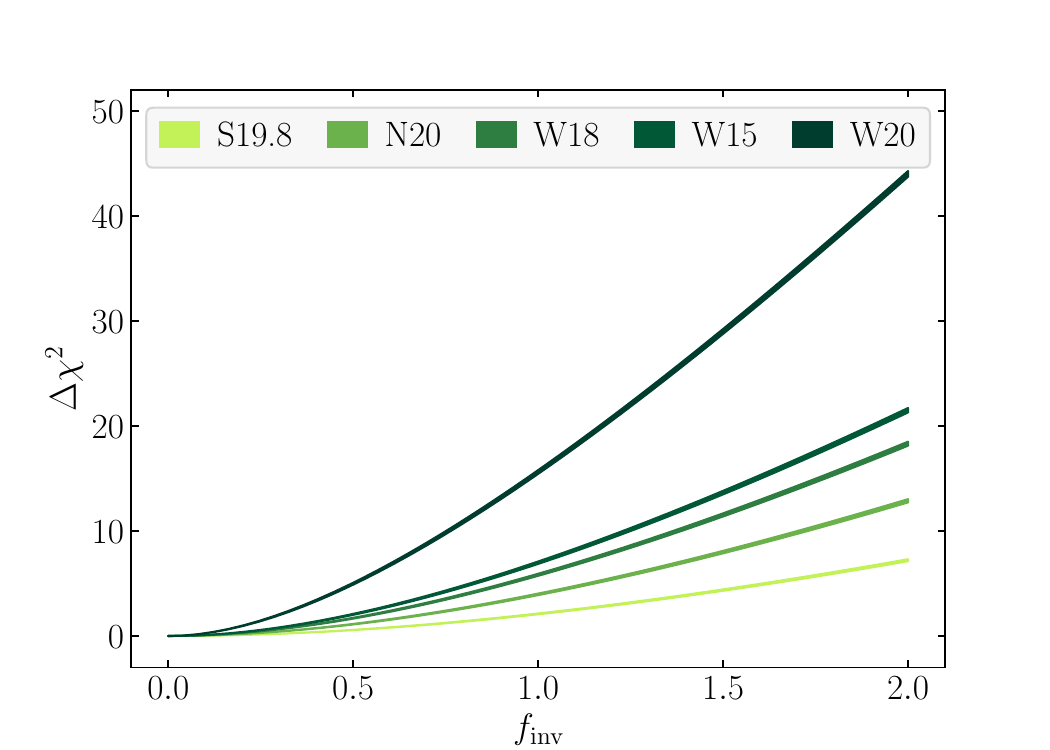}
    \hspace{-0.5cm}
    \includegraphics[width=0.5\linewidth]{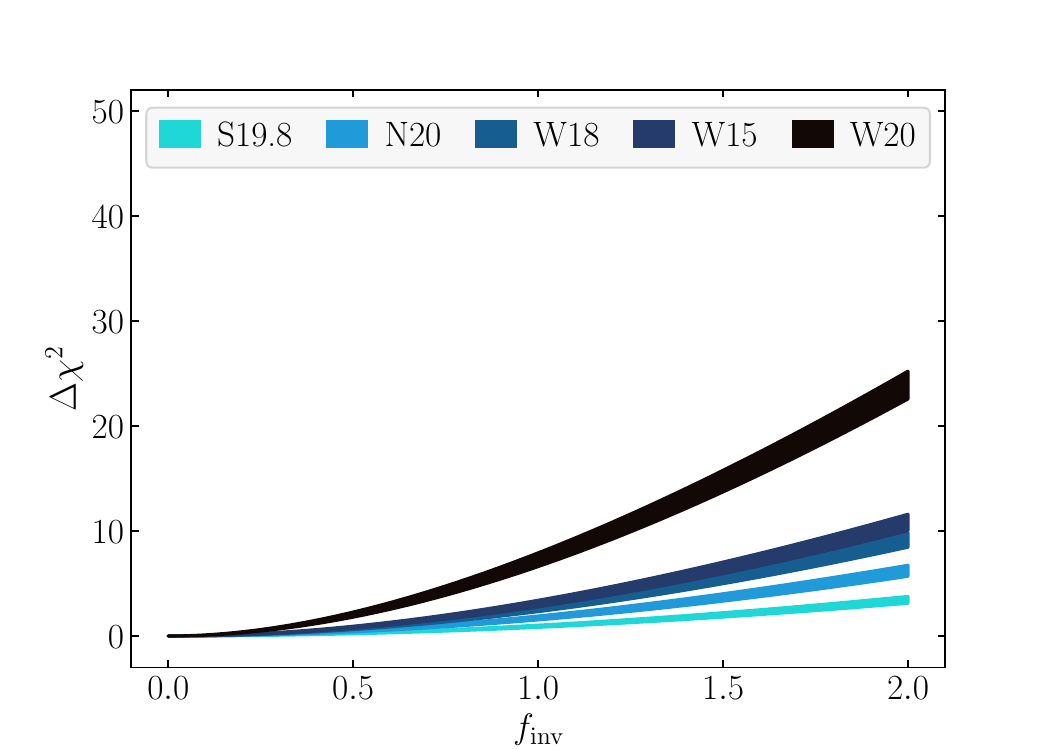}
    \caption{The $\Delta \chi^2$-profiles in the dependence of $f_{\mathrm{inv}}$ for 10 years of data with JUNO  (\textbf{Left}) and SK  (\textbf{Right}). The energy window is 12 - 30 MeV for both, hence \textbf{no} sidebands are used for SK. The width of the bands is given by the difference of the integrated (low) and binned (high) log-likelihood. For JUNO this difference is not visible.}
    \label{fig: WTchi2}
\end{figure}

In \cref{fig: WTchi2}, the $\chi^2$-profiles for \textbf{Test 1} with 10 years of data for JUNO  (\textbf{Left}) and SK (right) as a function of $f_{\mathrm{inv}}$. For the fiducial model W18, we can compare JUNO and SK which have a $\Delta \chi^2$ ($\chi^2-\chi_{\text{min}}^2$) of $\sim$6.1 and 3.0 at $f_\mathrm{inv}$=1, respectively.

The width of the bands in \cref{fig: WTchi2} shows the difference between the upper edge of the band (using spectral information -- binned) and the lower edge (not using spectral information -- integrated). For instance, in the right plot (SK) of \cref{fig: WTchi2}, for the weakest engine model (W20), we have a $\Delta \chi^2$ of $\sim$6.9 for the integrated analysis, and $\sim$8.2 for the binned analysis, at $f_\mathrm{inv}$=1.The energy window of interest is 12 - 30 MeV for both experiments. 

As expected, for reference models with higher $f_\mathrm{BH}$ we have a higher sensitivity to an invisible component, $f_{\mathrm{inv}}$, added to a purely 'visible' DSNB signal. Note that JUNO performs better due to the better signal-to-background ratio assumed. 

\begin{figure}[h!]
    \centering
    \includegraphics[width=0.5\linewidth]{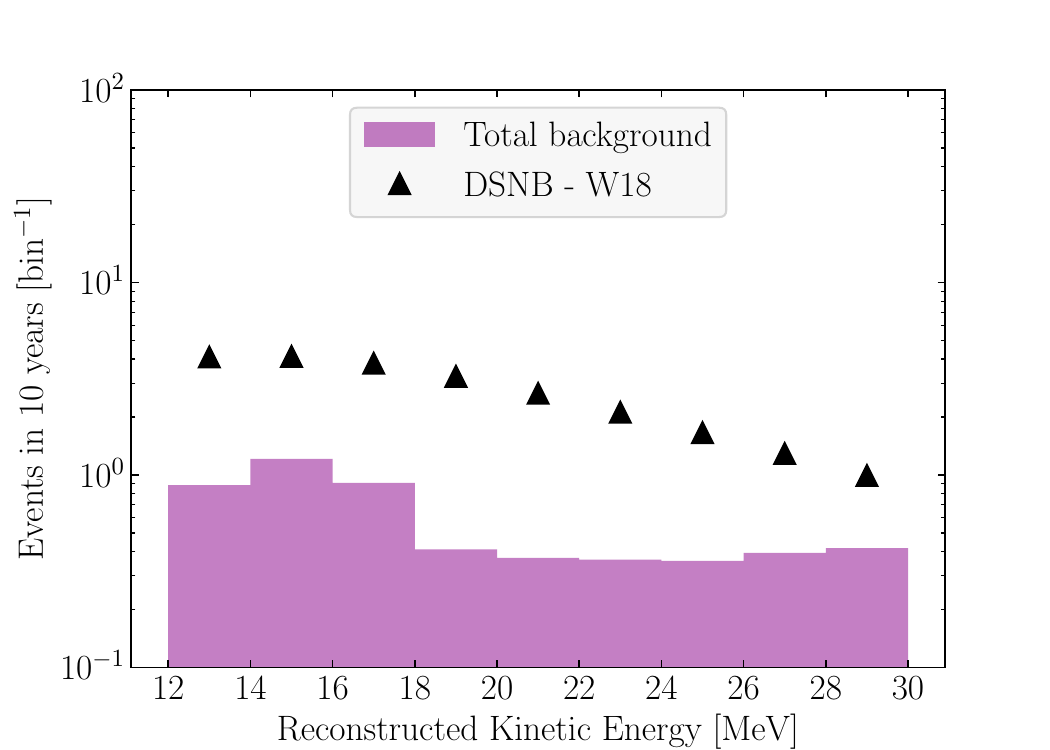}
    \hspace{-0.5cm}
    \includegraphics[width=0.5\linewidth]{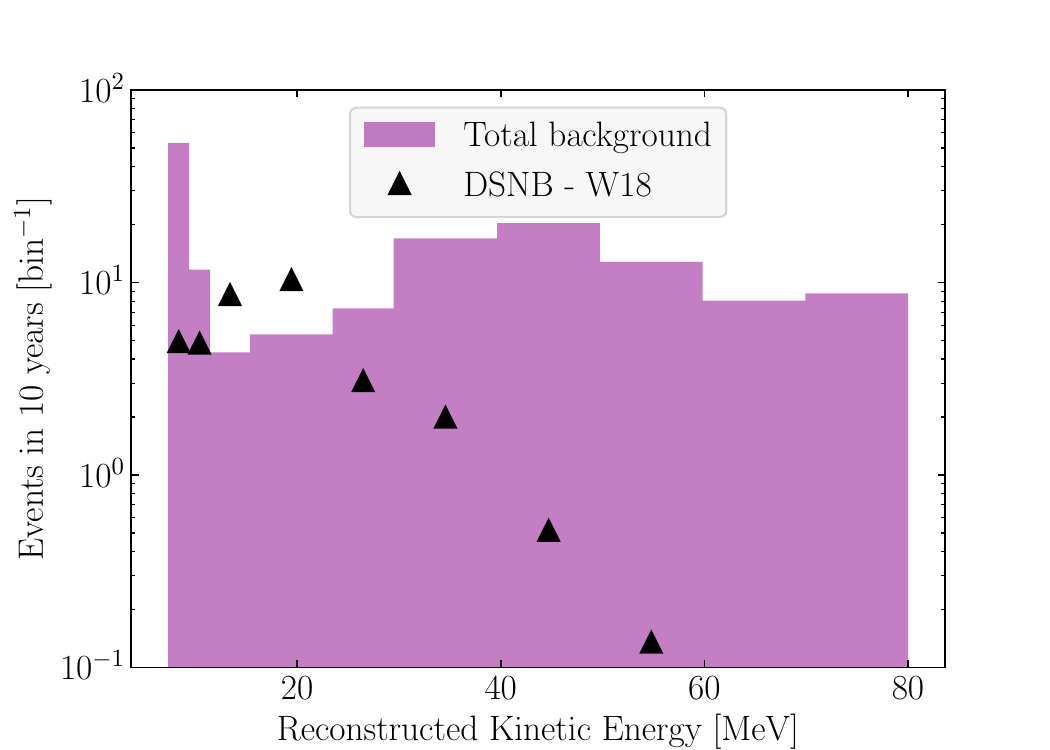}
    \caption{The signal and total background event numbers expected for 10 years of data with JUNO  (\textbf{Left}) and SK  (\textbf{Right}). While for JUNO the spectral shape for signal and background is similar, for SK it is significantly different.}
    \label{fig: sigAndBg}
\end{figure}

In \cref{fig: WTchi2}, the different band thickness of the two plots show that utilizing spectral information is much less important for JUNO than for SK due to the spectral shape of the backgrounds. In \cref{fig: sigAndBg}, the signal and background shapes are similar for JUNO while they differ for SK at high energies. Note that the binned approach has only very limited sensitivity to the spectral shape of the DSNB, which is instead mostly influencing the overall signal event rate due to the energy dependence of the IBD cross-section.

\paragraph{Test 2:} \textbf{\textit{What is the black hole fraction?}}\\

\begin{figure}[h!]
    \centering
    \includegraphics[width=0.5\linewidth]{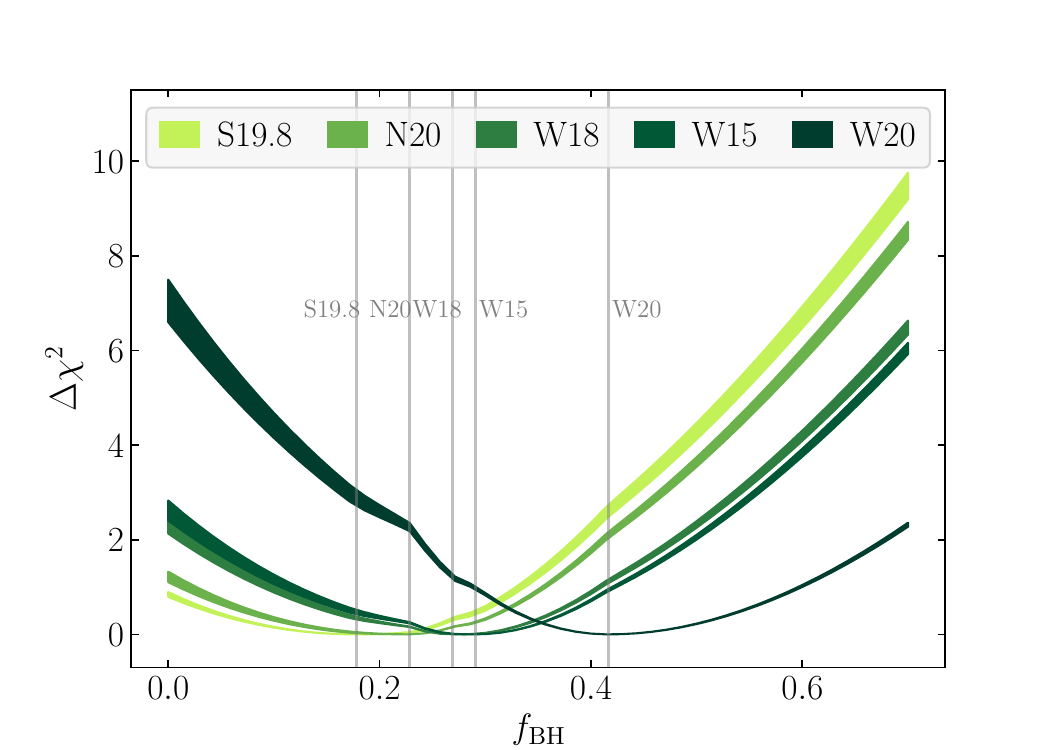}
    \hspace{-0.5cm}
    \includegraphics[width=0.5\linewidth]{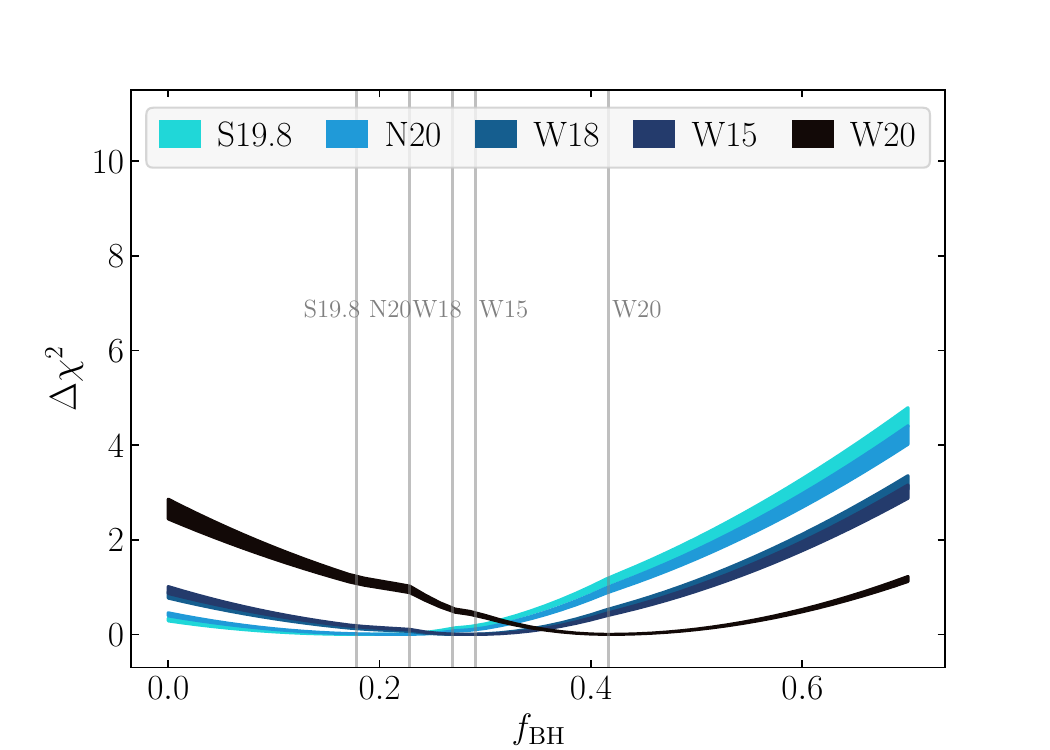}
    \caption{The $\Delta \chi^2$-profiles in dependence of $f_{\mathrm{BH}}$ for 10 years of data with JUNO  (\textbf{Left}) and SK  (\textbf{Right}). For both the energy window used is 12 - 30 MeV, note that \textbf{no} sidebands are used for SK. The profiles are not smooth, which is caused by the interpolation between the models (see \cref{fig:model1}). The bands indicate the difference of the integrated (low) and binned (high) log-likelihoods.}
    \label{fig: TTchi2}
\end{figure}

The $\chi^2$-profiles for \textbf{Test 2} are presented in \cref{fig: TTchi2} for 10 years of data with JUNO and SK. In this test, we compare each five engine models to an interpolated spectra, see \cref{sec:method}. Therefore, each profile has its minimum at the black hole-fraction of the reference model used. The irregularity in the profiles results from the interpolation, see \cref{fig:model1}. For reference, we compare JUNO and SK for the fiducial model (W18), which have a $\Delta \chi^2$ of $\sim$2.5 and $\sim$0.9 for $f_\mathrm{BH}$=0, respectively.

The upper edge of the bands corresponds to using spectral information and the lower edge to using an integrated spectrum. In contrast to \textbf{Test 1}, the difference between binned and integrated approach is visible for JUNO. In \textbf{Test 1}, the visible component is not altered, and the invisible component is scaled by $f_{\mathrm{inv}}$. However, in \textbf{Test 2}, both components are interpolated, so we compare different spectral shapes when scanning $f_{\mathrm{BH}}$. 

\section{Discussion}
\label{sec:discuss}
In this section, we discuss the results and their implications, focusing on limitations of the present study, and potential areas for future research. In particular, first we discuss how the background rejection for SK is controlled by the non-analysis region (sidebands), then we discuss how the sensitivity scales across the next years as other experiments begin operation. Finally, we highlight the dependence the neutrino spectral parameters. 
\subsection{Sidebands}
\label{sec:sideband}

\begin{figure}[h!]
    \centering
    \includegraphics[width=0.5\linewidth]{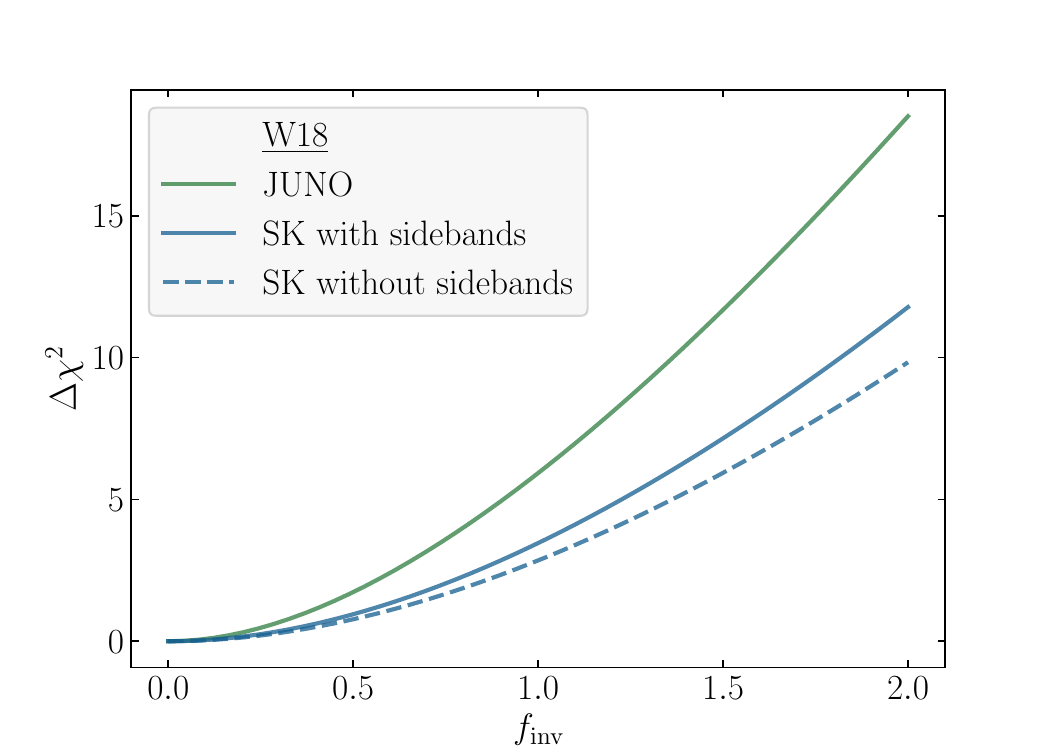}
    \hspace{-0.5cm}
    \includegraphics[width=0.5\linewidth]{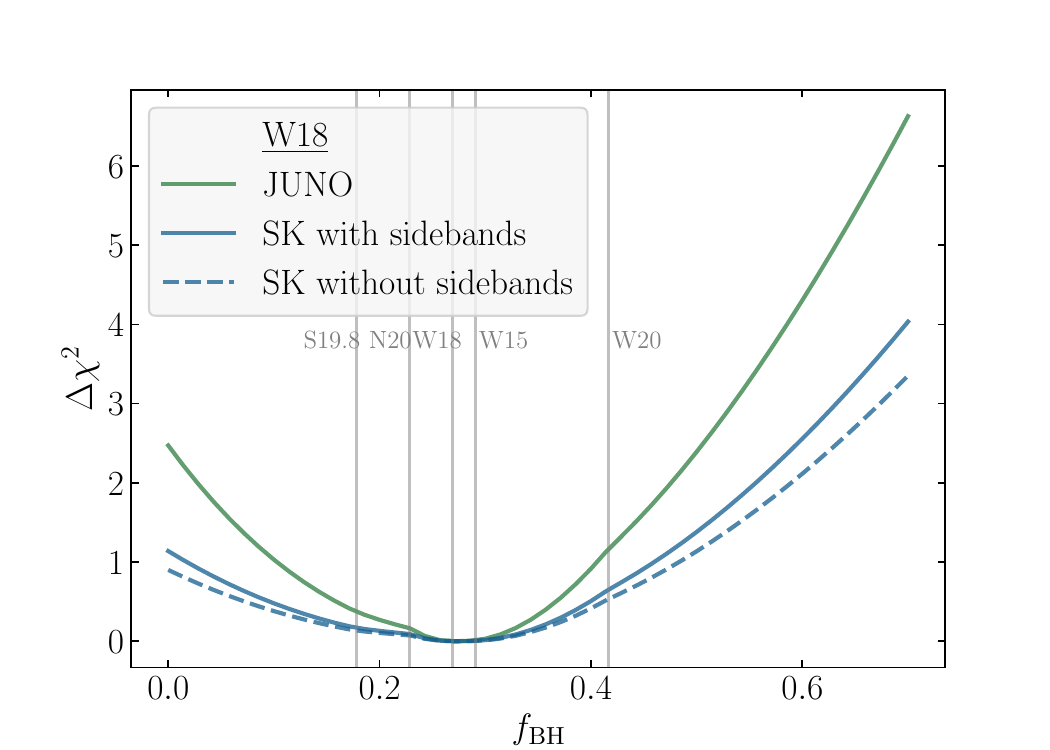}
    \caption{Comparison of the $\Delta \chi^2$-profiles of Test 1  (\textbf{Left}) and Test 2  (\textbf{Right}) for 10 years of data with JUNO and SK assuming the fiducial model W18. Here, the full energy region is used for SK (see \cref{fig:bg-spectra}), resulting in an increase in sensitivity compared to the analysis that is confined to the JUNO DSNB search window. For JUNO the energy region is still 12 - 30 MeV.}
    \label{fig: chi2Comp}  
\end{figure}

It is relevant to note that the plots in the previous section did not include 'sidebands', an important part of the SK analysis. In this section, we want to quantify the positive effect of including them. 
In \cref{fig: WTchi2,fig: TTchi2} we confined an the energy region for both experiments to the JUNO DSNB search window of 12 - 30 MeV. However, the sensitivity of the analyses using spectral information benefits from using the whole SK energy region. The energy windows outside the signal region is called a 'sideband'. Although sidebands do not give any signal information, they further constrain the backgrounds in the fit. Hence, we repeated both sensitivity tests with sidebands for SK. Note that this is only feasible when using spectral information, as the signal-to-background ratio is much lower in the integrated approach.

In \cref{fig: chi2Comp} the $\chi^2$-profiles of the analysis with the inclusion of sidebands for 10 years of data with SK are shown for \textbf{Test 1} (left panel) and \textbf{Test 2} (right panel) for the fiducial model W18. The corresponding JUNO profiles are also shown for comparison. While the sensitivities of SK improve when using the whole energy region (compare \cref{fig: WTchi2} and \cref{fig: TTchi2}), JUNO is still more sensitive to both $f_\mathrm{BH}$ and $f_{\mathrm{inv}}$,
\subsection{Near Future Prospects}
\label{sec:timeline}
\begin{figure}[b!]
    \centering
    \includegraphics[width=0.8\linewidth]{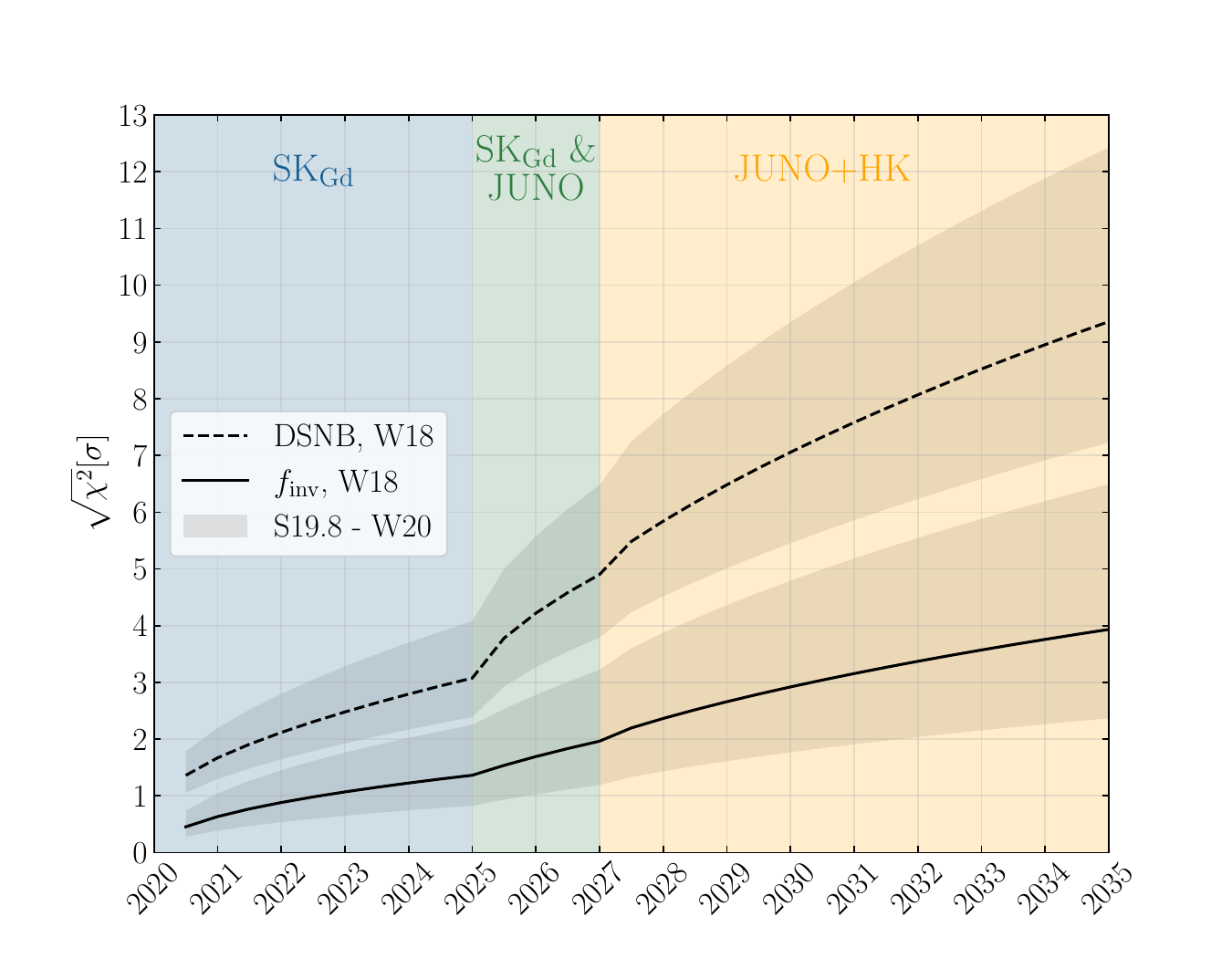}
    \caption{Development of the sensitivity to an invisible component by failed CCSNe in the DSNB (solid) as well as the DSNB itself (dashed) from 2020 - 2035. The different periods of detectors are indicated in colours. The lines correspond to the expected sensitivities for the fiducial model W18 ($f_{\mathrm{BH}}\sim0.27$) and the shaded area between the models S19.8 ($f_{\mathrm{BH}}\sim0.18$) and W20 ($f_{\mathrm{BH}}\sim0.42$).}
    \label{fig: timeline}
\end{figure}
Given the scarce statistics of a single detector, here we investigate how the sensitivity to an invisible component in the DSNB would evolve in the next years with a \textit{combined} measurement of all DSNB sensitive detectors. For this, we take 
\begin{align}
    \sqrt{\Delta \chi^2(f_{\mathrm{inv}}=1)}
    \label{eq:stat}
\end{align}

as our sensitivity measure---using the unchanged model for both NS and BH components. We calculate our sensitivity for SK, JUNO and HK individually and combine according to the expected operation schedule, see \cref{sec:exp}. We start our study with SK in 2020, the beginning of their Gadolinium-phase, the high efficiency DSNB search.

The resulting timeline is shown in \cref{fig: timeline} as a solid line for our fiducial model W18 and as a shaded area for the extreme models S19.8 and W20. We find that for the fiducial model W18 ($f_{\mathrm{BH}}\sim0.27$) a sensitivity to an invisible component in the DSNB of 3$\sigma$ can be reached in 2030, while 5$\sigma$ might be possible for the more extreme case of model W20, which is the weakest engine model with the largest $f_{\mathrm{BH}}\sim0.42$. A large increase is visible when JUNO begins operation.
    
To highlight the progression of sensitivity of $f_{\mathrm{inv}}$ in context of the overall DSNB search, we show the sensitivity to the DSNB detection assuming a pure background expectation, using our assumptions for signal, backgrounds, and detector performance. For this we apply
\begin{align}
    \chi^2(f_{\mathrm{inv}})\simeq-2\sum_{i} \ln L_i(f_{\mathrm{inv}})
    +\sum_j \left(\frac{\beta_j}{\sigma_{j}}\right)^2,
\end{align}
with the likelihood
\begin{align} 
L_i(f_{\mathrm{inv}}) = P\Bigl( s^{\mathrm{NS}}_{i}+\sum_j \beta_j b_{ji}+f_{\mathrm{inv}}s^{\mathrm{BH}}_{i},\quad \sum_j b_{ji}\Bigl),
\end{align}
using the same components as \cref{eq: likelihoodWT}, and we also present this in \cref{fig: timeline}

The resulting sensitivity timeline is also shown in \cref{fig: timeline} with a dashed line, with the models unchanged and a shaded area to indicate the extreme models.  
\subsection{Varying neutrino spectral parameters}
\label{sec:vary}

\begin{figure}[b!]
    \centering
    \includegraphics[width=0.5\linewidth]{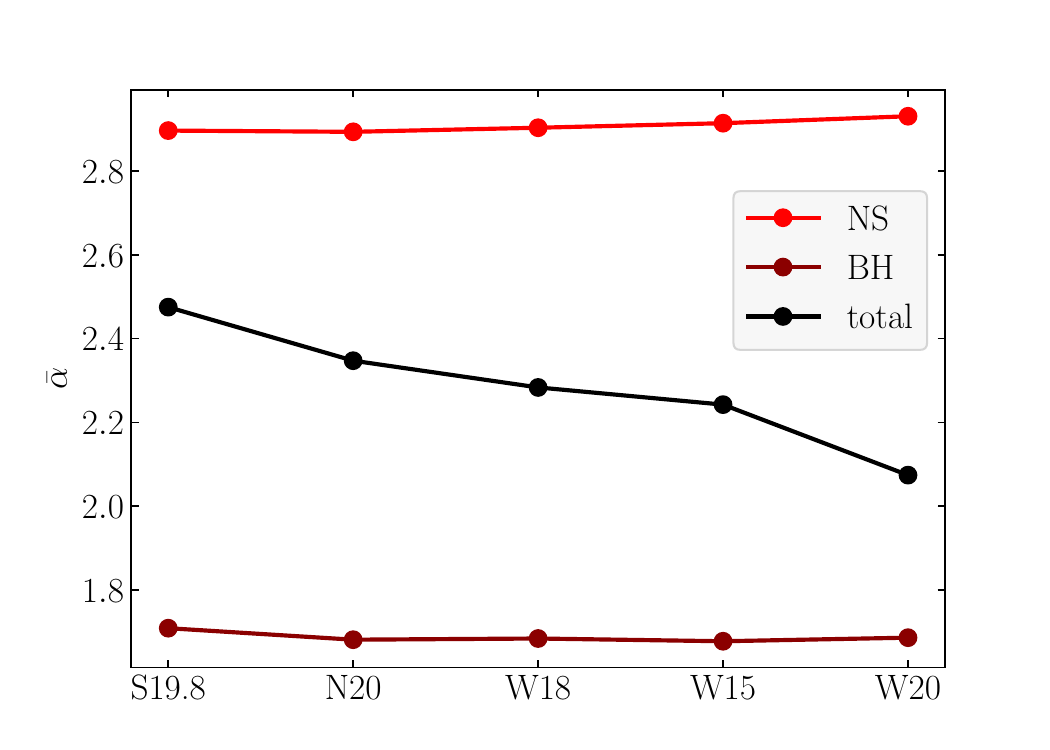}
    \hspace{-0.5cm}
    \includegraphics[width=0.5\linewidth]{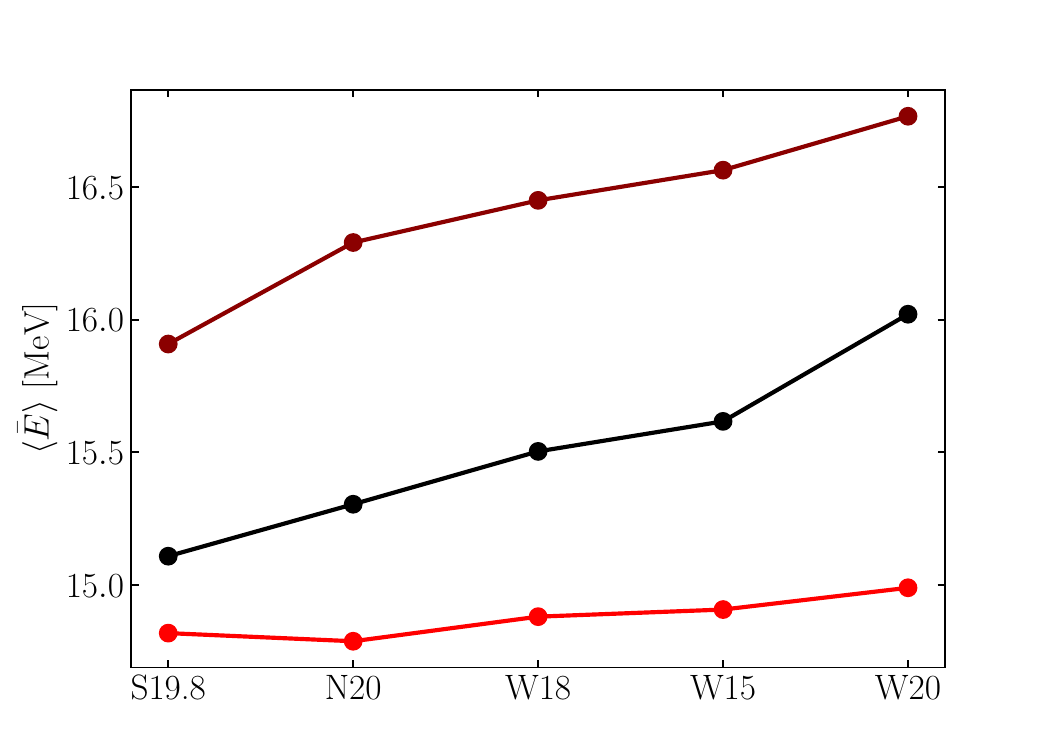}
    \caption{Spectral shape or \textit{pinching} parameter, $\alpha$, (\textbf{Left}) and mean energy, $\langle E \rangle$, (\textbf{Right}) of the IMF-averaged, time-integrated neutrino spectra for each supernova engine model \cite{Sukhbold2016, Kresse2021}}%
    \label{fig:params}
\end{figure}


The overall normalization and spectral parameters are predicted to be clarified with future sky surveys \cite{Lien2010} and gravitational wave observations \cite{Ecker2025}, respectively, for the visible component. For the invisible component, the dominant uncertainty is the overall normalization, which is the dominant effect in this work. However, in this section we vary the spectral parameters of the invisible component, $\alpha_{\mathrm{BH}}$ and $\langle E \rangle_{\mathrm{BH}}$, and carry out \textbf{Test 1} to see the effect on the final sensitivity.

In this discussion, we will refer to the 'instantaneous' spectral parameters ${\alpha}$ and ${\langle E \rangle}$, which are the input parameters for the 200 reference models used in the simulation \cite{Garching}. We refer the reader to \cite{Kresse2021} for the full discussion of parameters used in the simulation. After IMF-weighting and integrating over time, these sets of models can be described by the spectral parameters of this time-integrated, IMF-averaged flux, which we denote $\bar{\alpha}$ and $\bar{\langle E \rangle}$, respectively.

First, we present the effect of shifting $\alpha_\textrm{BH} $, the instantaneous pinching parameter, from 1 - 3. These models are provided by Kresse \textit{et al} \cite{Kresse2021}, and are available in addition to the models in the rest of the work which have $\alpha_\textrm{BH} = 2.0$ \cite{Garching}. 

\begin{figure}[h!]
    \centering
    \includegraphics[width=0.5\linewidth]{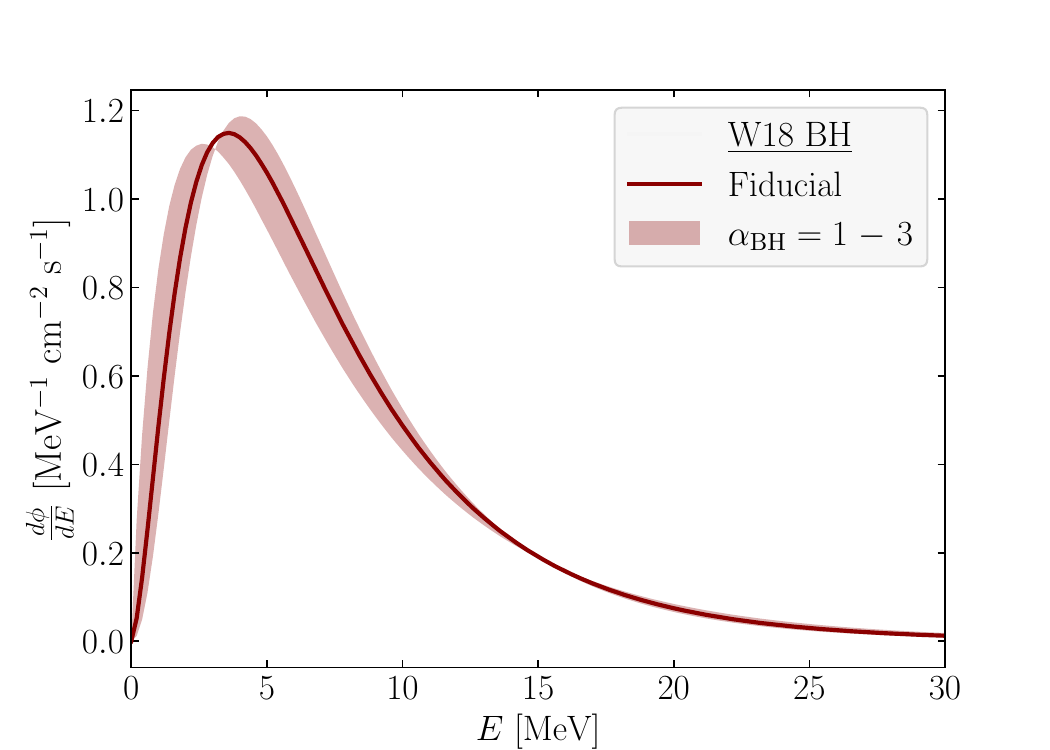}
    \hspace{-0.5cm}
    \includegraphics[width=0.5\linewidth]{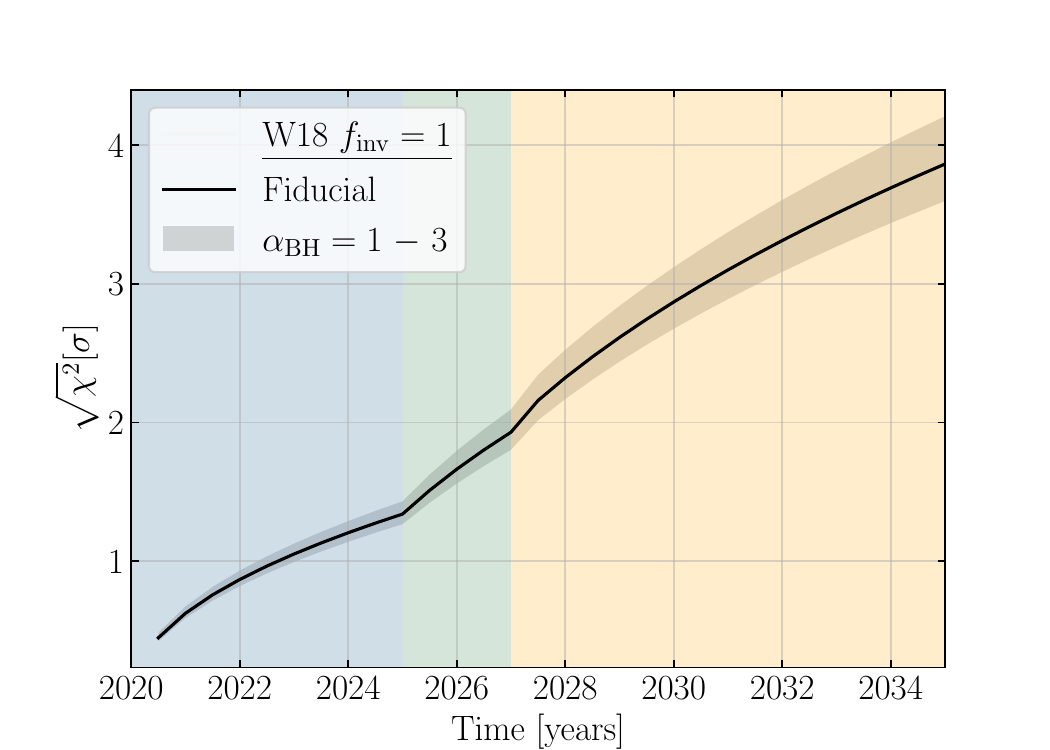}
        \caption{Plots of varying the spectral shape pinching parameter, $\alpha_\textrm{BH}$, using the models provided \cite{Garching} from ${\alpha_\textrm{BH}}=$1 - 3, note the fiducial has ${\alpha_\textrm{BH}}=2$. \textbf{Left}: Plot of the flux as a function of energy for the invisible black hole-forming component. \textbf{Right}: Sensitivity to $f_{\mathrm{inv}}$ as a function of time. The colours indicate the periods in which different experiments are running (SK-Gd only, SK-Gd and JUNO, JUNO and HK).}
    \label{fig:alpha}
\end{figure}

In \cref{fig:alpha}, we show \textbf{Test 1}, where the NS component is fixed, but the total black hole fraction is varied. We fix $f_{\mathrm{inv}}$=1 (\cref{eq:stat}), and change the spectra by varying ${\alpha}$ using the models provided \cite{Garching}. It is clear from \cref{fig:alpha} that changing the pinching parameter of the invisible component does not have a large impact on the sensitivity of detecting the black hole-forming component.

In \cref{fig:params}, notice that the mean energy of the invisible component of time-integrated neutrino spectra, $\bar{\langle E \rangle}_{\textrm{BH}}$, is larger for weaker engine models. With a weaker engine there are more collapses with a longer period before black hole-formation, leading to a harder neutrino spectra. To estimate the effect on the sensitivity we vary $\bar{\langle E \rangle}_{\textrm{BH}}$ from 15 - 18.5 MeV. The variation in mean energy is also linked to the value of maximum baryonic neutron star mass, $M_{\textrm{lim}}$, which can also be extracted from a gravitational wave signal. Kresse \textit{et al} \cite{Kresse2021} fix 2.7$M_\odot$ for the nominal models, but also make available models with 2.3$M_\odot$ and 3.1$M_\odot$ \cite{Garching}.

\begin{figure}[h!]
    \centering
    \includegraphics[width=0.5\linewidth]{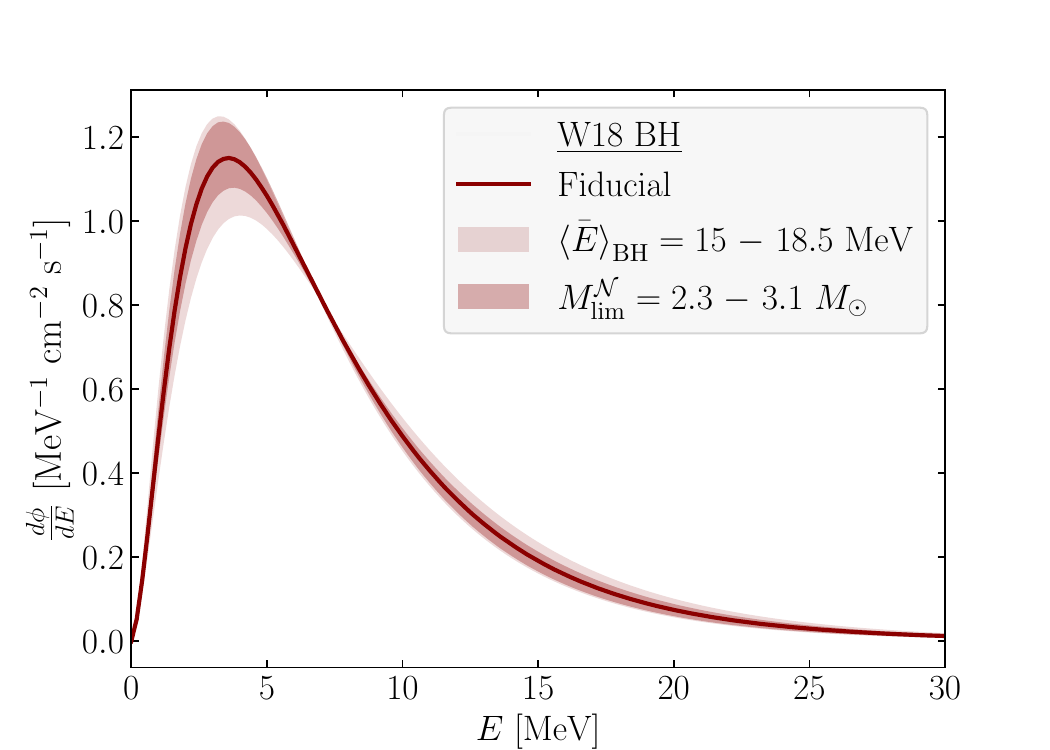}
    \hspace{-0.5cm}
    \includegraphics[width=0.5\linewidth]{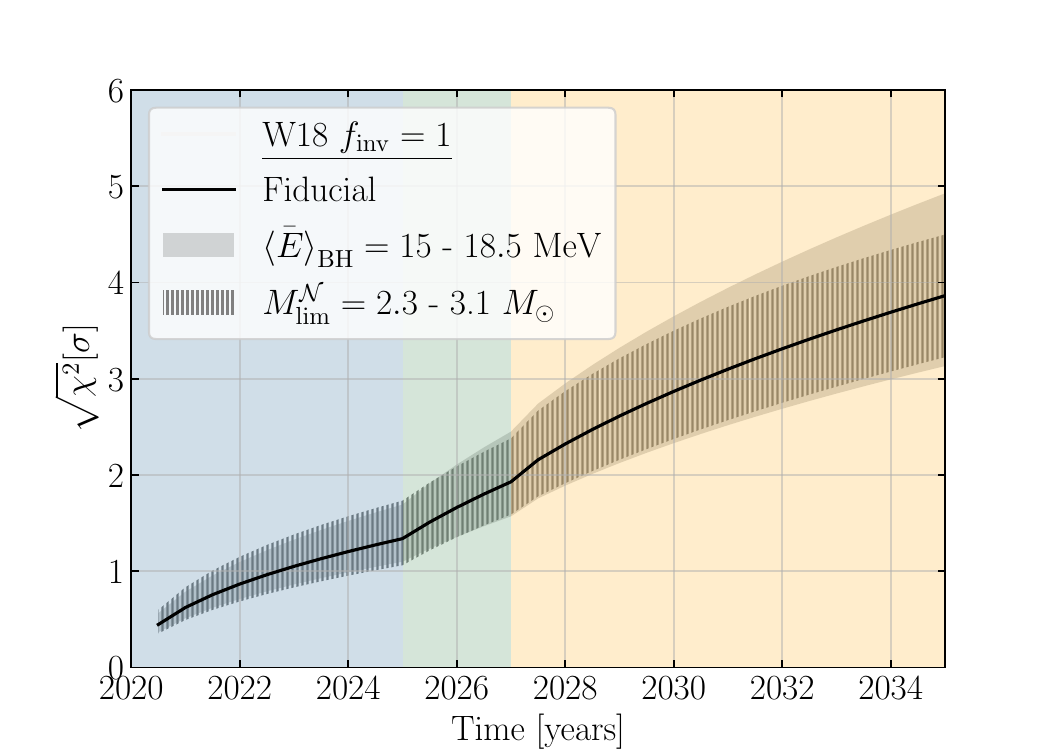}
    \caption{Plots of varying the mean energy $\bar{\langle E \rangle}_{\textrm{BH}}$ and baryonic neutron star mass limit $M_{\textrm{lim}}$. We present the effect of varying $\bar{\langle E \rangle}_{\textrm{BH}}$ using the DSNB parameterization the  and varying $M_{\textrm{lim}}$ with the models provided \cite{Garching}, but scaling them to have the same normalization as the fiducial model ($M^{\mathcal{N}}_{\textrm{lim}}$). Note the fiducial model has 2.7$M_\odot$ and $\sim16.5$ MeV. \textbf{Left}: Plot of the flux as a function of energy for the invisible black hole-forming component. \textbf{Right}: Sensitivity to $f_{\mathrm{inv}}$ as a function of time. The colours indicate the periods in which different experiments are running (SK-Gd only, SK-Gd and JUNO, JUNO and HK).}
    \label{fig:mean}
\end{figure}

In \cref{fig:mean} we show a plot varying $\bar{\langle E \rangle}_{\textrm{BH}}$---fitting the spectrum and then recreating it using the parameterization described in \cref{sec:theory}. To emphasize the degeneracy with the neutron star baryonic mass limit, we also show $2.3M_\odot$ and $3.1M_\odot$, scaled to have the same normalization as the fiducial model ($2.7M_\odot$) to isolate the dependence of the spectral parameters on $M^{\mathcal{N}}_{\textrm{lim}}$. Notice how the variation in sensitivity is larger for $M^{\mathcal{N}}_{\textrm{lim}}$ for more than the first five years, and then the variation in $\bar{\langle E \rangle}_{\textrm{BH}}$ becomes larger.

\section{Conclusion}
\label{sec:conc}
Standing on the threshold of large neutrino observatories discovering the DSNB, it is clear that there will be many degeneracies in the observed low-statistics spectra. We will have to rely on co-operation with other disciplines to extract underlying features from the signal. First DSNB observations will be extremely timely with new data from breakthrough optical telescope projects and gravitational wave detectors. They are likely to provide precise predictions for the neutron star-forming, or visible, component of the DSNB. Contrariwise, neutrino detectors have the potential to provide us information on the black hole-forming, or invisible, component not accessible for telescopes and interferometers.  

While the interpretation of these initial DSNB measurements will be complicated by theoretical and astrophysical degeneracies, we have shown in this paper that early DSNB observations possess the potential to constrain both the existence, and fraction, of invisible supernovae. Our findings also reveal that the detectability and capability to characterize the invisible supernova component depend critically on the normalization and spectral parameters of the visible and invisible components of the DSNB. In the case of a 'median' model that we consider to be representative for today's span of available explosion engines, DSNB observations will be able to reach a 3$\sigma$ detection of the black hole forming component within 10 years. 

We have demonstrated that the DSNB can probe one of the most elusive populations in stellar evolution, offering new insight into the death of massive stars. We hope this work also charts a way forward---even with the first DSNB events, we will have access to physics that has so far been out of reach.

\begin{acknowledgments}
We would like to thank Hans-Thomas Janka and Daniel Kresse for in-depth discussions throughout the course of this work. We would also like to thank Alberto Garfagnini, Yu-Feng Li, Hans-Thomas Janka and Daniel Kresse for carefully proofreading the manuscript. We are grateful to the Mainz Institute for Theoretical Physics (MITP) for hospitality and partial support during the Workshop \cite{MITP2024DSNB}. This work has been supported by Cluster of Excellence PRISMA${}^{+}$ (EXC 2118/1-39083149) funded by the DFG within the German Excellence strategy. Generation of the DSNB model data at Garching was partially supported through the CRC \textit{Neutrinos and Dark Matter in Astro- and Particle Physics} (SFB-1258-283604770), and Cluster of Excellence ORIGINS (EXC-2094-390783311). 
\end{acknowledgments}

\bibliographystyle{unsrt}
\bibliography{main}

\section{Appendix}
\begin{figure}[hb!]
    \centering
    \includegraphics[width=0.5\linewidth]{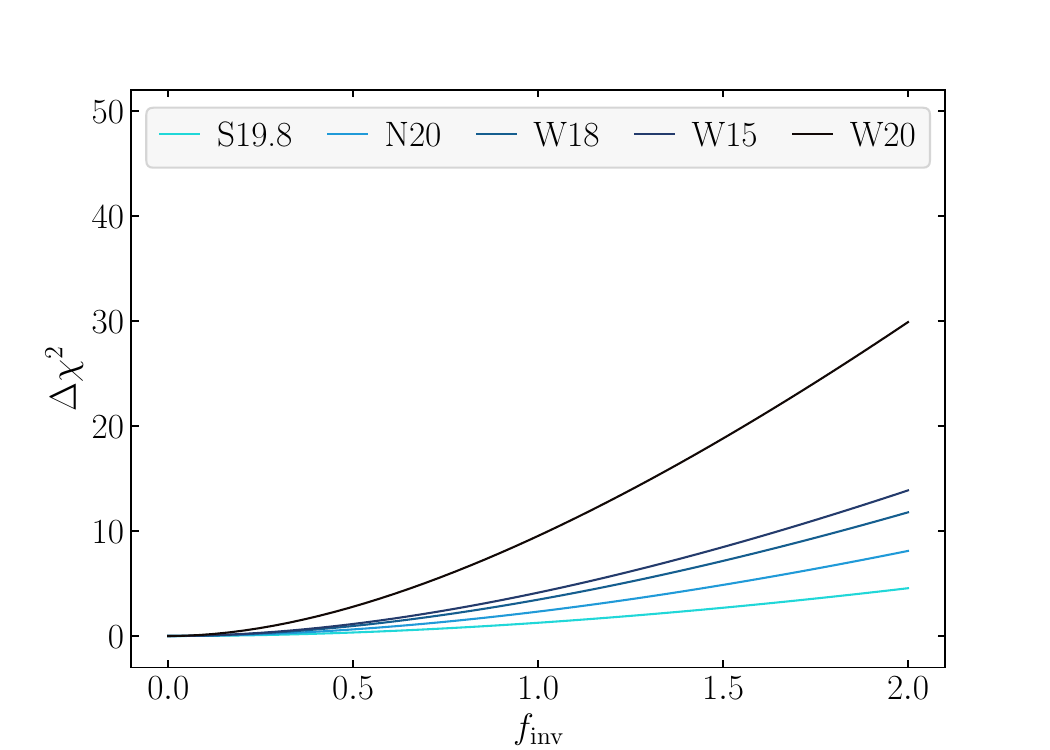}
    \hspace{-0.5cm}
    \includegraphics[width=0.5\linewidth]{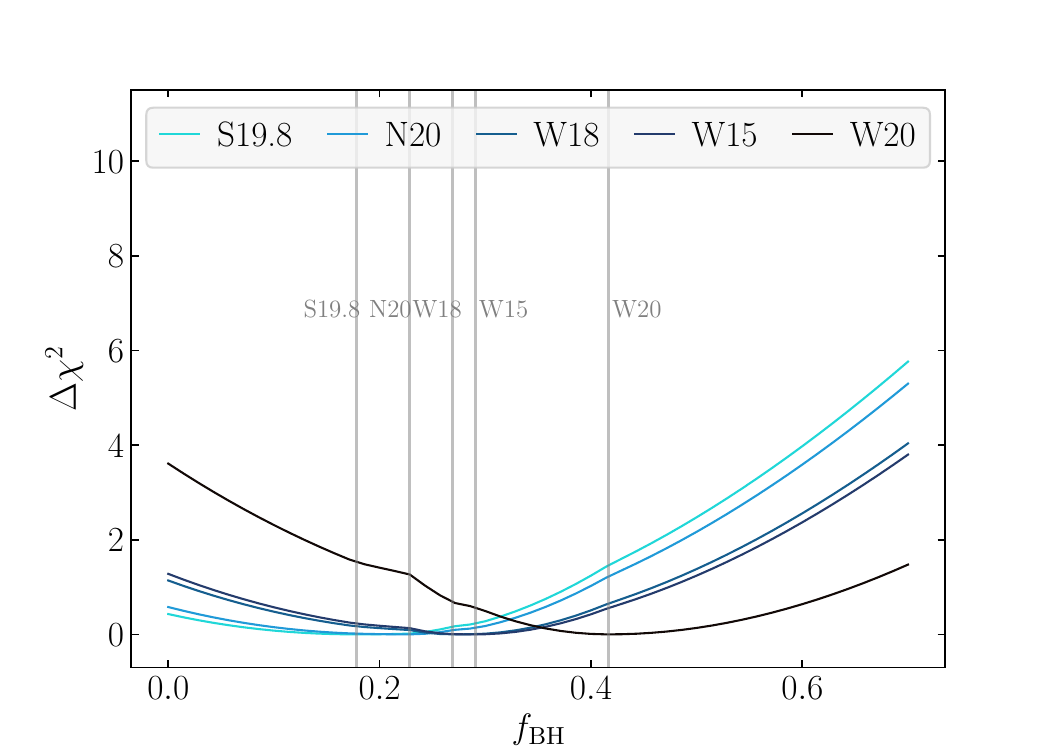}
    \caption{The $\chi^2$-profile of both \textbf{Test 1} and \textbf{Test 2} 10 years of data with SK. Here, sidebands \textit{are} used.}
    \label{fig: chi2SK_SB}
\end{figure}

\end{document}